\documentclass[10pt,twocolumn,nofootinbib,superscriptaddress,showpacs,showkeys,preprintnumbers,floatfix,byrevtex]{revtex4-1}

\usepackage{soul}

\setstcolor{red}	                                                                       
\usepackage[normalem]{ulem}  							

\usepackage{epsfig}
\usepackage{amsmath,amssymb}
\usepackage{bm}
\usepackage{multirow}
\usepackage{graphicx}
\usepackage{ulem}
\usepackage[usenames,dvipsnames,svgnames,table]{xcolor}

\newcommand{\eqn}[1]{Eq.\,\eqref{#1}}

\newcommand{\keV}{\,\mathrm{keV}}
\newcommand{\MeV}{\,\mathrm{MeV}}
\newcommand{\GeV}{\,\mathrm{GeV}}

\begin{document}

\title{Effects of sterile neutrino and extra-dimension on big bang nucleosynthesis}

\author{Dukjae Jang}\email{havevirtue@ssu.ac.kr}
\affiliation{Department of Physics and OMEG Institute, Soongsil
University, Seoul 156-743, Korea}

\author{Motohiko Kusakabe}\email{mkusakab@nd.edu}
\affiliation{Center for Astrophysics, Department of Physics, University of Notre Dame, Notre Dame, IN 46556, USA}

\author{Myung-Ki Cheoun}\email{cheoun@ssu.ac.kr (Corresponding Author)}
\affiliation{Department of Physics, Soongsil University, Seoul 156-743, Korea}

\date{\today}
\begin{abstract}
By assuming the existence of extra-dimensional sterile neutrinos in big bang nucleosynthesis (BBN) epoch, we investigate the sterile neutrino ($\nu_{\rm s}$) effects on the BBN and constrain some parameters associated with the  $\nu_{\rm s}$ properties. First, for cosmic expansion rate, we take into account effects of a five-dimensional bulk and intrinsic tension of the brane embedded in the bulk, and constrain a key parameter of the extra dimension by using the observational element abundances. Second, effects of the $\nu_{\rm s}$ traveling on or off the brane are considered. In this model, the effective mixing angle between a $\nu_{\rm s}$ and an active neutrino depends on energy, which may give rise to a resonance effect on the mixing angle. Consequently, reaction rate of the $\nu_{\rm s}$ can be drastically changed during the cosmic evolution. We estimated abundances and temperature of the $\nu_{\rm s}$ by solving the rate equation as a function of temperature until the sterile neutrino decoupling. We then find that the relic abundance of the $\nu_{\rm s}$ is drastically enhanced by the extra-dimension and maximized for a characteristic resonance energy $E_{\rm res}\gtrsim 0.01$ GeV. Finally, some constraints related to the $\nu_{\rm s}$, mixing angle and mass difference, are discussed in detail with the comparison of our BBN calculations corrected by the extra-dimensional $\nu_{\rm s}$ to observational data on light element abundances.
\end{abstract}

\maketitle

\section{Introduction}

Over the past few decades, a considerable number of studies has been conducted on the neutrino oscillation with a great success of measuring neutrino mixing angles. But, some experiments for the neutrino oscillation revealed more or less disagreements with the three-flavor neutrino model, which termed as the neutrino anomalies, as reported in LSND\,\cite{Aguilar:2001ty}, MiniBoone\,\cite{AguilarArevalo:2010wv}, reactor experiments\,\cite{Bhattacharya:2011ah} and gallium experiments\,\cite{Giunti:2010zu}. One of the approaches for explaining the neutrino anomalies is to presume the existence of the hypothetical fourth neutrino, which is called as sterile neutrino, because the sterile neutrino does not interact with other particles excepting through a mixing with active neutrinos.

Very recently, the IceCube experiment reported a new constrained region for the parameter space of the mixing angle and the mass-squared differences for the $1\,\mathrm{eV}$ mass scale sterile neutrino \cite{TheIceCube:2016oqi}, in which the parameter space by previous LSND and MiniBoone data are largely excluded. But, if we recollect that $1\,\mathrm{keV}$ cosmological sterile neutrino is still under discussion for a dark matter candidate and the relic neutrino search is being considered, it would be an interesting discussion to consider effects of the sterile neutrino in the big bang nucleosynthesis (BBN) epoch and deduce related parameters from the observational data with the comparison to the IceCube experimental data analysis.

Among many scenarios of the sterile neutrino, P\"as {\it et al.} \cite{Pas:2005rb} assumed that the sterile neutrino is a gauge-singlet particle and can travel on or off our 3 + 1 dimensional brane embedded in a large extra dimension bulk similarly to the graviton in the brane-world cosmology. According to the cosmology, ordinary matter fields are confined to a three-dimensional space in the high dimensional bulk. Originally, the brane-world cosmology was suggested to explain the hierarchy problem, the large scale difference between the standard model force and the gravity\,\cite{ArkaniHamed:1998rs,Shiu:1998pa}. Randall and Sundrum suggested a new solution of the hierarchy problem by introducing noncompact extra dimensions\,\cite{Randall:1999ee,Randall:1999vf}. Ref. \cite{Pas:2005rb} suggested a model, in which a sterile neutrino can propagate in the bulk and brane similarly to the graviton.  They derived a new formula of resonant active-sterile neutrino oscillation and found an allowed region of the resonance energy from the comparison to available experimental data.

If the production rate of this kind of sterile neutrino is always smaller than the cosmic expansion rate, the abundance of the sterile neutrino never reaches the equilibrium value.  The effect of the sterile neutrino on BBN is then completely negligible.  This situation has been considered recently \cite{Aeikens:2016rep}, and a parameter region where the sterile neutrino abundance is extremely small has been searched by an analytical estimate.  However, as shown in this paper,  observational constraints on primordial abundances do not exclude the situation that the sterile neutrino is abundantly produced in thermal bath and its abundance attains the equilibrium value in the early universe.  Furthermore, the observational abundance of $^4$He is possibly explained by the effect of the sterile neutrino better than in standard BBN model, as argued in this paper.

In this study, we adopt the same scheme as Ref. \cite{Pas:2005rb} and study effects of a sterile neutrino in an extra-dimensional universe by a numerical BBN calculation in detail. Especially, we considered not only the matter effects but also wave packet formalism to describe the oscillation between active and sterile neutrinos in the five-dimensional universe. Since the primordial element abundances can be measured with a good precision by the recent great advent of astrophysical spectroscopic observations, the BBN study turns out to be a useful test bed for deriving the cosmological constraints on nonstandard models.
For example, some parameters in the modified gravity models, such as $f(R)$ and $f(G)$ gravity, were constrained in detail \cite{Kusakabe:2015ida}. In addition, effects of some supersymmetric (SUSY) particles in the early universe have been investigated and parameters, i.e., the lifetime and mass, can be constrained \cite{Kusa14}.

We include effects of the extra dimensional sterile neutrino in the BBN epoch as follows. The cosmic expansion rate is modified by the large extra dimension \cite{Ichiki:2002eh, Sasankan:2016ixg} as well as the energy density \cite{Lee:1977ua, Sato:1977ye} of the sterile neutrino traveling on or off our 3 + 1 dimensional brane. Then the modified Friedmann equation and the energy density of decoupled sterile neutrino may change the primordial element abundances. Therefore, the parameters relevant to the extra dimension and the sterile neutrino can be constrained by using observational data of primordial light element abundances.

This paper is composed as follows. In section \,\ref{2}, we briefly review the sterile-active neutrino oscillation in the extra-dimension sterile neutrino model and address how to describe the evolution for the number abundance of sterile neutrino in the early universe in the model. In section \,\ref{3}, results of primordial nuclear abundances by the model are presented. From the results, in Sec.\,\ref{4}, we discuss the constrained parameter region from the comparison of BBN calculation results to observational abundance data. Section\,\ref{5} contains a summary and conclusions of this article. We derive the flavor change probability of the sterile neutrino in the current extra-dimension model in Appendix \ref{appendix1}. A result of solving the Boltzmann equation for the sterile neutrino and its comparison with that of the rate equation are shown in Appendix \ref{appendix2}.

\section{Theoretical Model}
\label{2}

We presume that the universe is five-dimensional and the sterile neutrino travels on or off the five-dimensional space as in Ref. \cite{Pas:2005rb}. We simply consider only one sterile neutrino and assume that sterile neutrinos interact with matter particles only via its mixing with an active neutrino. The decay of the sterile neutrino is not considered in this model.

\subsection{Modified cosmic expansion rate from extra-dimension}\label{sec2a}
According to Ref. \cite{Binetruy:1999hy}, the cosmic expansion rate in a five-dimensional universe is given by,
\begin{eqnarray}
\frac{\dot{a}_0^2}{a_0^2} && =\frac{\kappa^2}{6}\rho_B+\frac{\kappa^4}{36}\rho_b^2+\frac{\mathcal{E}}{a_0^4}-\frac{K}{a_0^2}~, \nonumber \\
&& =\frac{\kappa^2}{6}\rho_B+\frac{\kappa^4}{36}\rho_\Lambda^2+\frac{\kappa^4}{18}\rho_\Lambda\rho+\frac{\kappa^4}{36}\rho^2+\frac{\mathcal{E}}{a_0^4}-\frac{K}{a_0^2}~,
\label{eq1}
\end{eqnarray}
where $a_0$ is the scale factor for the four-dimensional space time. $\rho_B$ denotes the bulk energy density in the universe. Energy density of the brane, $\rho_b$, is given as a sum of ordinary energy density ($\rho$) and energy density ($\rho_\Lambda$) stemming from the intrinsic tension on the brane, $\rho_b = \rho + \rho_{\Lambda}$. $\mathcal{E}$ is an integration constant. The five-dimensional analogue of the gravitational constant, $G_{(5)}$, is related to the five-dimensional Planck mass, $M_{(5)}$ and the constant $\kappa$, as follows
\begin{align}
\kappa^2=8 {\pi} G_{(5)} = M^{-3}_{(5)}~.
\label{eq2}
\end{align}
The last term in the right-hand side of Eq. (\ref{eq1}) vanishes in the flat universe where the curvature constant is $K=0$. We can choose $\rho_\Lambda$ by following Refs.\,\cite{Randall:1999ee, Randall:1999vf}
\begin{align}
\frac{\kappa^2}{6}\rho_B+\frac{\kappa^4}{36}\rho^2_\Lambda=0 ~.
\end{align}
Then, the cosmic expansion rate of the standard cosmology is recovered for $\rho \ll \rho_{\Lambda}$ by the identification \cite{Csaki:1999jh,Cline:1999ts} of
\begin{align}
8\pi G \simeq \frac{\kappa^4\rho_\Lambda}{6} ~,
\end{align}
where
$G$ is Newton's constant.
Our final expansion rate was obtained as
\begin{align}
\frac{\dot{a}^2}{a^2} \simeq \frac{8{\pi}G}{3}\rho+\frac{\mathcal{E}}{{a}^4} ~,
\label{eq5}
\end{align}
where the index 0 in the scale factor on the brane $a_0$ has been omitted.

The first term of the right-hand side in Eq.\,\eqref{eq5} is identical with the cosmic expansion rate in the standard cosmological model. Note that only the second term comes from the effect of the extra dimension. The free parameter $\mathcal{E}$, which is a kind of an integration constant in the five-dimensional Einstein equation, affects the primordial abundances \cite{Ichiki:2002eh,Sasankan:2016ixg}.  The initial temperature of our BBN calculation is $T_9 =T/(10^9~{\rm K}) =100$ with $T$ the temperature.  We then take the value of the second term at $T_9 =100$, i.e., $\mathcal{E}/a_{\rm i}^4$ with $a_{\rm i}$ the scale factor at the initial temperature, as a parameter.

\begin{figure}
\centering
\includegraphics[width=7.5cm]{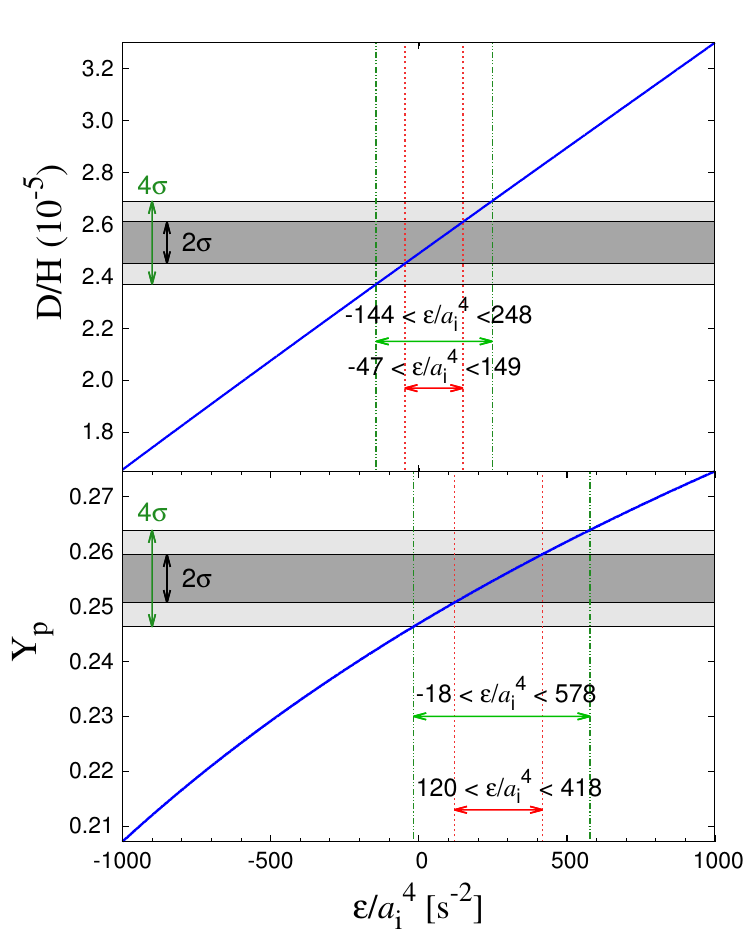}
\caption{(Color online) Deuterium (the top panel) and $\mathrm{^4He}$ (the bottom panel) abundances as a function of $\mathcal{E}/a_{\rm i}^4$. Shaded and dark-shaded regions are 4 $\sigma$ and 2 $\sigma$ ranges, respectively, for the observational primordial abundances. We adopt the observational value of D/H=${(2.53\pm 0.04)\times 10^{-5}}$ \cite{7} and $Y_{\rm p}=(0.2551\pm0.0022)$ \cite{8}. From the observational data, the $\mathcal{E}$ value is constrained as $120<\mathcal{E}/a_{\rm i}^4~({\rm s}^{-2})<149$ (2 $\sigma$) and $-18<\mathcal{E}/a_{\rm i}^4~({\rm s}^{-2})<248$ (4 $\sigma$).}
\label{fig1}
\end{figure}

Figure \ref{fig1} shows calculated abundances of the deuterium (number ratio of D/H) and $^4$He (the mass fraction $Y_{\rm p}$) as a function of $\mathcal{E}/a_{\rm i}^4$.  The abundances of D and $^4$He are monotonically increasing with the increase of $\mathcal{E}/a_{\rm i}^4$.
The standard BBN model corresponds to the case of $\mathcal{E}=0$.  In this case, the predicted deuterium abundance is within the observational $2\sigma$ limit, while the $^4$He abundance is out of  the $2\sigma$ limit.  When the nonzero value of $\mathcal{E}$ is considered, we find the $2\sigma$ allowed region from the both D and $^4$He abundances in the region of $120<\mathcal{E}/a_{\rm i}^4~({\rm s}^{-2})<149$.

When the $\mathcal{E}$ value is increased, the cosmic expansion rate is also increased.  Since the cosmic time scale for a fixed temperature is shorter, the neutron to proton ratio at the $^4$He synthesis is larger.  As a result, the $^4$He abundance after the BBN is larger.  In the late time of BBN, the deuterium is effectively destroyed by the reactions $^2$H($d$,$n$)$^3$He and $^2$H($d$,$p$)$^3$H.  The shorter cosmic time scale leads to the earlier freeze-out of the destruction reactions. Subsequently, the larger final deuterium abundance is obtained.

\subsection{Relic abundance of sterile neutrino}\label{sec2b}
Since we include the sterile neutrino which has a finite mass $m_{\nu_{\rm s}}$, its energy density $\rho_{\nu_{\rm s}}$ is added to the ordinary density in \eqn{eq5}.  The total energy density is thus changed as,
\begin{align}
\rho=\rho_{\rm standard} + \rho_{\nu_{\rm s}}~,
\label{eq6}
\end{align}
where the first and the second terms indicate energy densities of standard model particles and sterile neutrinos, respectively. The value of $\rho_{\nu_{\rm s}}$ is roughly evaluated as,
\begin{align}
\rho_{\nu_{\rm s}} \simeq \begin{cases} n_{\nu_{\rm s}} \left\langle E_{\nu_{\rm s}}\right\rangle \ (\mathrm{for \ relativistic \ case})\\
 n_{\nu_{\rm s}} m_{\nu_{\rm s}}  \ \quad (\mathrm{for \ non\mbox{-}relativistic\ case}).
\end{cases}
\label{eq7}
\end{align}
Here $n_{\nu_{\rm s}}$ and $\left\langle E_{\nu_{\rm s}}\right\rangle$ are the number density and the averaged energy of the sterile neutrino, respectively. The energy density is separated into relativistic and non-relativistic cases which depend on temperature $T_{\nu_{\rm s}}$ and $m_{\nu_{\rm s}}$. Namely, for $m_{\nu_{\rm s}}>\left\langle E_{\nu_{\rm s}}\right\rangle \sim 3T_{\nu_{\rm s}}$, the sterile neutrino is non-relativistic. Otherwise, it is relativistic. Since the $n_{\nu_{\rm s}}$ in \eqn{eq7} is a key quantity to determine the energy density of the sterile neutrino, we calculate the number density of the sterile neutrino in the following way.

In the hot early universe, the sterile neutrino can stay in an equilibrium state when its production rate is enough large. However, with the decrease of temperature, the sterile neutrino is decoupled from the equilibrium state. The decoupling condition is that the reaction rate of the sterile neutrino $\Gamma_{\nu_{\rm s}}$ becomes smaller than the cosmic expansion rate $H \equiv \dot{a}/a$. At that time, the ratio $Y_{\nu_{\rm s}}$ between $n_{\nu_{\rm s}}$ and the entropy density $s$ freezes out, i.e., does not change (see Chap. 5 of Ref. \cite{Kolb1994}). In order to describe the $Y_{\nu_{\rm s}}$ evolution, we exploit the following rate equation,
\begin{align}
\frac{x}{Y_{\rm EQ}}\frac{d Y_{\nu_{\rm s}}}{dx}=-\frac{\Gamma_{\nu_{\rm s}}}{H} \left[\left(\frac{Y_{\nu_{\rm s}}}{Y_{\rm EQ}}\right)^2-1 \right],
\label{eq8}
\end{align}
where $x \equiv m_{\nu_{\rm s}}$/$T$ and $Y_{\rm EQ}=n_{\rm EQ}/s$ is a ratio of the equilibrium number density to the entropy density in the co-moving unit volume given in terms of temperature $T$,
\begin{align}
s=\frac{2\pi^2}{45}g_{*\mathrm{S}}T^3~,
\end{align}
where $g_{*\mathrm{S}}$ is defined in terms of the degrees of freedom of particle $i$,
\begin{align}
g_{*\mathrm{S}}=\Sigma_{i={\rm boson}}\,g_i \left(\frac{T_i}{T} \right)^3+\frac{7}{8}\,\Sigma_{i={\rm fermion}}\,g_i \left(\frac{T_i}{T} \right)^3.
\label{eq9}
\end{align}
We note that the temperature of the sterile neutrino is the same as that of thermal bath, i.e., $T_{\nu_{\rm s}} =T$, until the decoupling of the sterile neutrino.

From the assumption that the sterile neutrino interacts with other particles via only mixing, the production rate of the sterile neutrino $\Gamma_{\nu_{\rm s}}$ is given by a product of the probability of the flavor change of $\nu_a \leftrightarrow \nu_{\rm s}$ via mixing, $P_{\rm as}$, and averaged weak interaction rate $\left\langle \Gamma_{\rm weak} \right\rangle$ \cite{Barbieri:1989ti,Barbieri:1991,Enqvist:1991qj},
\begin{align}
\Gamma_{\nu_{\rm s}}= P_{\rm as} \left\langle \Gamma_{\rm weak} \right\rangle.
\label{eq11}
\end{align}
 In this study, we adopt the simplest case in which one sterile neutrino mixes with only one active neutrino \cite{Pas:2005rb}, and assume that the tau neutrino has the mixing for simplicity.  We then adopt the average weak interaction rate of $\nu_\tau$, i.e., $\left\langle \Gamma_{\rm weak} \right\rangle \rightarrow \Gamma_{\tau}=2.9\,G^2_{\rm F} T^5$  \cite{Enqvist:1991qj}, where $G_F$ is the Fermi constant. We set the initial condition $Y_{\rm i}=0$ in \,\eqn{eq8}. Because the reaction rate of sterile neutrino depends on parameters, the sterile neutrino does not always stay in equilibrium at the initial time within all parameter space.  This is in contrast to active neutrinos that are consistently in equilibrium well before BBN.  The distribution function of the sterile neutrino is then not always the equilibrium function.  For large reaction rates relative to the cosmic expansion rate, the equilibrium abundance realizes quickly, while for small reaction rates, the abundance remains much smaller than $Y_{\rm EQ}$.  This is the reason why we assume that the initial abundance of the sterile neutrino is equal to zero.

In addition, in \eqn{eq8}, we neglect the effect of an extra-dimension on the cosmic expansion rate. As shown in Sec.\,\ref{sec2a}, observations of light element abundances strongly constrain the value of $\mathcal{E}/a_{\rm i}^4$. The cosmic expansion rate in the early epoch until the sterile neutrino decoupling is, therefore, not allowed to deviate significantly from that in the standard model. The rate equation is then not affected significantly.

\subsection{Modified flavor-change probability}\label{sec2c}

In solving the rate equation, flavor-change probability in \eqn{eq11} should account for the extra-dimensional and matter effects.  Since the trajectories of sterile neutrinos in the bulk and active neutrinos on the brane are different, their flavor-change probability is different from that in free space \cite{Pas:2005rb}. In addition, the neutrino oscillation is affected by the matter effect. These two effects can be treated similarly to the effective potential in the Mikheev-Smirnov-Wolfenstein (MSW) physics \cite{Wolfenstein:1977ue,Mikheev:1986wj}. When the matter effect \cite{Notzold:1988, Barbieri:1989ti, Dolgov:2001} and difference of geodesic are included, the effective mixing angle is derived as 

\begin{align}
  {\sin^2}{2\tilde{\theta}}=\frac{\sin^2 2\theta}{Q_\alpha^2(\theta, \delta m^2, E_{\rm res}; T, E)},
\label{eq15}
\end{align}
where
  we defined a parameter $Q_\alpha$ for the modification of the mixing angle given by
\begin{widetext}
\begin{equation}
  Q_\alpha(\theta, \delta m^2, E_{\rm res}; T, E) =
  \sqrt{\mathstrut
    \sin^2 2\theta +\cos^2 2\theta \left [
    1 +\frac{C_\alpha G_{\rm F}^2 T^4 E^2}{\cos 2 \theta \alpha \delta m^2}
    - \left( \frac{E}{E_{\rm res}} \right)^2 \right]^2
  },
  \label{eq_add1}
\end{equation}
\end{widetext}
where $\alpha$ is the fine structure constant, $C_{e}=1.22$ (for $\nu_e$) and $C_{\mu, \tau}=0.34$ (for $\nu_\mu$ and $\nu_\tau$) are flavor ($\alpha$) dependent constants. We used $C_{\tau}=0.34$ because we considered only $\nu_\tau$. $\theta$ is the bare mixing angle between the sterile and active neutrinos, $\delta m^2$ denotes the mass squared difference, and $E$ is the energy of the sterile neutrino. The resonance energy $E_{\rm{res}}$ is given \cite{Pas:2005rb} by
\begin{align}
E_{\rm res} = \sqrt{\frac{\delta m^2 \cos2\theta}{2 \epsilon_{\rm s}}},
\label{eq16}
\end{align}
where
  $\epsilon_{\rm s}=(D_b-D_B)/D_b$ is a shortcut parameter describing the fractional difference between the geodesic in the bulk $D_B$ and that on the brane $D_b$.

We assume that the sterile neutrino is relativistic before the decoupling, and use the value of $E=3.151\,T_{\nu_{\rm s}}$, which is the averaged energy for the relativistic fermion with ${T_{\nu_{\rm s}}}$ the temperature of the sterile neutrino.

The probability of the flavor change of sterile and active neutrinos is derived from the wave packet treatment \cite{{Giunti:1991}, {Fukugita:2010}} as
\begin{eqnarray}
P_{as} \approx \begin{cases} \sin^2 2 \tilde{\theta} \sin^2 \left( \frac{\delta m^2_{\rm{mat}} t_{\rm{sc}}}{4E} \right) & (\mbox{for}\ T \geq T_{\rm{eq}}) \\
\frac{1}{2} \sin^2 2 \tilde{\theta} & (\mbox{for}\ T \leq T_{\rm{eq}}),
\end{cases}
\label{eq17}
\end{eqnarray}
where we defined the effective mass-squared difference in matter, i.e.,
\begin{equation}
  \delta m^2_{\rm mat} = \delta m^2 Q_\alpha(\theta, \delta m^2, E_{\rm res}; T, E),
  \label{eq_add2}
\end{equation}
  and the scattering time scale of active neutrino
\begin{equation}
  t_{\rm sc} \simeq \frac{1}{G_{\rm F}^2 T^5}.
  \label{eq_add3}
\end{equation}
The typical temperature $T_{\rm eq}$ is defined related to the flavor-change probability as
\begin{eqnarray}
  T_{\rm eq} &=&\left( \frac{\delta m^2}{G_{\rm F}^2} \right)^{1/6} \nonumber \\
  &=& 44~{\rm MeV} \left(\frac{\delta m^2}{1~{\rm eV}^2} \right)^{1/6}.
  \label{eq_add4}
\end{eqnarray}
At this temperature, the scattering time scale of active neutrino and the overlap time scale of neutrino wave packets equal.  In addition, the matter effect becomes negligible somewhat below this equality temperature.

A formulation of the flavor-change probability including Eqs. (\ref{eq15}), (\ref{eq_add1}), (\ref{eq17})--(\ref{eq_add4}) is shown in Appendix \ref{appendix1}. By using Eqs. (\ref{eq8}), (\ref{eq11}), and (\ref{eq17}), the abundance of the sterile neutrino is calculated.

\section{Result of the rate equation}
\label{3}
By using the modified mixing probability, we solve the rate equation in the temperature interval of 100 GeV $\ge T \ge$ 1 MeV. This rate equation is approximation of the Boltzmann equation. The comparison of results of Boltzmann and rate equations is described in Appendix B. Figure\,\ref{fig2} shows the contours for the final values of $Y_{\nu_{\rm s}}$ calculated by the rate equation as a function of $\theta$ and $E_{\rm res}$.  For this figure and Figs. 3--5 in this section, the mass of sterile neutrino is taken to be $m_{\nu_{\rm s}} \approx (\delta m^2)^{1/2}=1$ eV for example.  We discuss the result from three viewpoints.
\begin{figure}
\centering
\includegraphics[width=8.5cm]{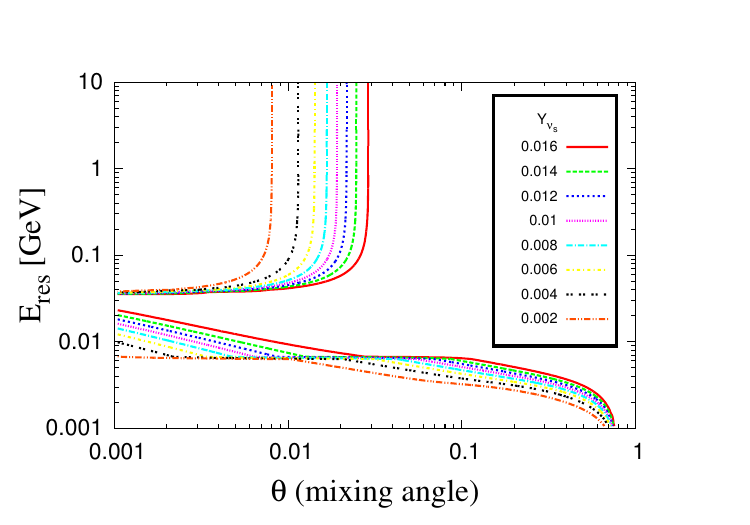}
\caption{(Color online) Contours for the final value of $Y_{\nu_{\rm s}}$ as a function of $\theta$ and $E_{\rm res}$ [GeV] for $m_{\nu_{\rm s}} =1$ eV.}
\label{fig2}
\end{figure}

\subsection{Mixing angle vs. $Y_{\nu_{\rm s}}$}
First, we explain the results in Fig.\,\ref{fig2} as a function of mixing angle for a given $E_{\mathrm{res}}$. The value of $Y_{\nu_{\rm s}}$ is larger for larger mixing angle.  As seen in Eqs. (\ref{eq11}) and (\ref{eq17}), the larger mixing angle produces the larger flavor-change probability and reaction rate.

\begin{figure}
\centering
\includegraphics[width=7.5cm]{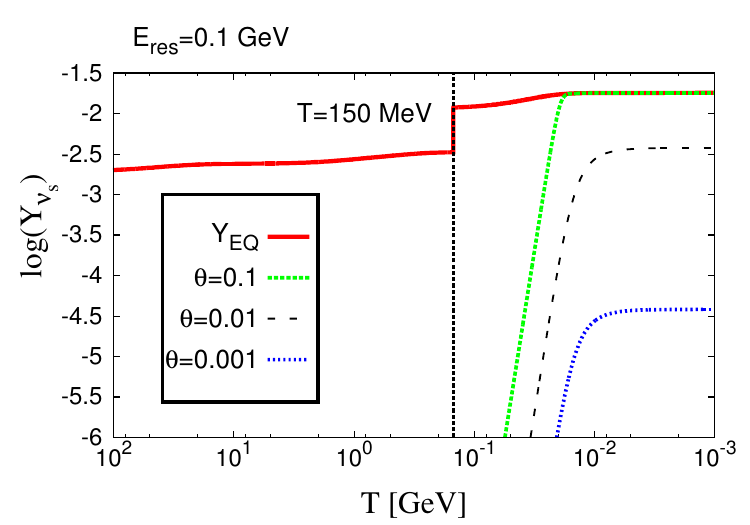}
\caption{(Color online) Temperature evolution of the abundance of sterile neutrino. The red-solid line indicates the equilibrium abundance $Y_{\rm{EQ}}$ and other lines show the abundance of the sterile neutrino $Y_{\nu_{\rm s}}$ for the mixing angle $\theta=0.1$ (higher dashed line), 0.01 (lower dashed line), and 0.001 (dotted line), respectively. The mass is $m_{\nu_{\rm s}} =1$ eV and the resonance energy is fixed as $E_{\rm res} =0.1\,\GeV$.  The black vertical dashed line at $T=150\,\MeV$ shows the temperature of the quark-hadron transition.}
\label{fig3}
\end{figure}

  Figure \ref{fig3} shows the temperature evolution of the abundance of sterile neutrino derived by solving the rate equation. The red-solid line indicates the equilibrium abundance $Y_{\rm{EQ}}$ and other lines show the abundance of the sterile neutrino $Y_{\nu_{\rm s}}$ for $\theta=0.1$ (higher dashed line), 0.01 (lower dashed line), and 0.001 (dotted line), respectively.  The resonance energy is fixed as $E_{\rm res} =0.1\,\GeV$ for example.  The black vertical dashed line at $T=150\,\MeV$ shows the temperature of the quark-hadron transition.

The equilibrium abundance $Y_{\rm EQ}$ is increased with decreasing $T$ because the number of degrees of freedom $g_{*S}$ decreases. This behavior is remarkably contrary to the decrease of $Y_{\nu_{\rm s}}$ by the exponential decrease of the equilibrium number density $n_{\rm EQ}$ after the sterile neutrino becomes non-relativistic. Because of the hadronization of the quark-gluon plasma (QGP) the abundance $Y_{\nu_{\rm s}}$ is increased with decreasing temperature in this epoch. This feature can be seen at the vertical line.  For this figure, the entropy density is calculated by a standard method \cite{Kolb1994} described in Ref. \cite{Ishida:2014wqa}

No sterile neutrino exists at initial time by the assumption. When the temperature decreases to $T\sim T_{\rm eq} =44$ MeV, however, the effective mixing angle increases (Sec. \ref{sec2c}). The production rate of the sterile neutrino then becomes large, and its number density approaches to the equilibrium line. In this parameter set, no resonance of the effective mixing angle occurs as the universe evolves. This can be understood by noting that the square bracket in Eq. (\ref{eq_add1}) is always close to or larger than unity. The temperature where the second and the third terms in the square bracket equal is given by $T_{\rm res,1} =11.2$ MeV [Eq. (\ref{eq_b65})]. At this temperature, the factor $(E/E_{\rm res})^2$ is significantly smaller than 1. Therefore, no resonance occurs in this model (see Appendices \ref{appendix1} and \ref{appendix2} for details on the resonant mixing in the early universe). After the matter term becomes negligible in Eq. (\ref{eq_add1}), the effective mixing angle is close to the bare mixing angle, i.e., $\sin^2 2 \tilde{\theta} \approx \sin^2 2 \theta$. Since the reaction rate is proportional to $\tilde{\theta}^2$, the final abundance is almost proportional to $\theta^2$ for small $\theta$ values (see curves of $\theta =0.01$ and $0.001$). For large $\theta$ values, the equilibrium abundance is realized before the decoupling of the sterile neutrino (the case of $\theta=0.1$). Fig.\,\ref{fig3} shows that the small mixing angle gives low abundance of the sterile neutrino for this parameter set.

\subsection{The resonance energy vs. $Y_{\nu_{\rm s}}$}
\begin{figure}
\centering
\includegraphics[width=7.5cm]{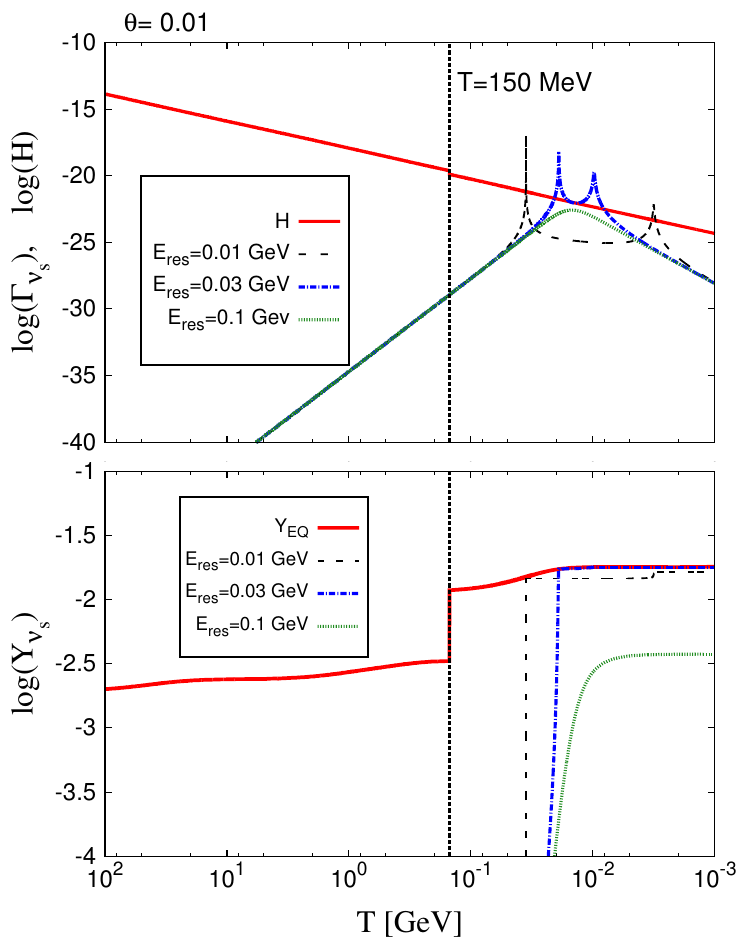}
\caption{(Color online) Temperature evolution of the reaction rate and the cosmic expansion rate (upper panel) and the abundance (lower panel). The mass is $m_{\nu_{\rm s}} =1$ eV and the mixing angle is fixed as $\theta=0.01$. The red-solid line indicates the cosmic expansion rate (upper panel) and the equilibrium abundances $Y_{\rm{\rm EQ}}$ (lower panel). Other lines show the abundance of the sterile neutrino $Y_{\nu_{\rm s}}$ for the resonance energy $E_{\rm{res}} = 0.01\,\rm{GeV}$ (long dashed line), $0.03\,\rm{GeV}$ (dashed-dotted line) and $0.1\,\rm{GeV}$ (dotted line).}
\label{fig4}
\end{figure}

Second, the resonance energy dependence is interpreted similarly, because it is related to the reaction rate of the sterile neutrino. Figure \ref{fig4} shows the temperature evolution of reaction rate $\Gamma_{\nu_{\rm s}}$ and $H$ (upper panel) and $Y_{\nu_{\rm s}}$ (lower panel). The value of mixing angle $\theta$ is fixed as 0.01. The black vertical dashed line at $T=150\,\rm{MeV}$ corresponds to the temperature of the quark-hadron transition. At the temperature, the equilibrium abundance of the sterile neutrino is increased because of decreasing $g_{*\mathrm{S}}$.

  The reaction rate is very small in the high temperature region, $T \gtrsim T_{\rm eq} =44$ MeV since a large matter term hinders the effective mixing angle [Eqs. (\ref{eq15}) and (\ref{eq_add1})]. For the cases of $E_{\rm res} =0.01$ and $0.03$ GeV, maximal enhancements of the effective mixing angle occur twice, respectively. At the peaks, the production rate of the sterile neutrino is larger than the cosmic expansion rate. As a result, the abundance of sterile neutrino approaches the equilibrium abundance.  However, resonances occur for short periods [cf. Eqs. (\ref{eq_b16}), (\ref{eq_b17}) and (\ref{eq_dis_3})], and the equilibrium abundance is not reached.  At the first resonance, the second term of the square brackets in Eq. (\ref{eq15}) cancels the third term, and the effective mixing angle increases. At the second resonance, the first term, i.e., unity, cancels the third term, and the effective mixing angle increases again (See Appendix \ref{appendix2_6} for details). At these resonances, the abundance of the sterile neutrino suddenly increases (the lower panel). The abundance is flat excepting the resonance epochs because the sterile neutrino is decoupled from the equilibrium.

  For the case of $E_{\rm res} =0.1$ GeV, there is no resonance of the effective mixing angle, as explained for Fig. \ref{fig3}. Therefore, the reaction rate does not have a peak, and the abundance evolves smoothly.

  As seen in Fig. \ref{fig4}, the abundance of sterile neutrino is significantly enhanced by the extra-dimensional correction to the effective mixing angle. For large values of $E_{\rm res} \gtrsim 0.04$ GeV, no resonance in the effective mixing angle appears along the cosmic evolution. This parameter region asymptotes to the standard model of four dimensional universe. For small values of $E_{\rm res} \lesssim 0.04$ GeV, resonances appear in the mixing angle, and the final sterile neutrino abundance is enhanced. We observe that the final abundance is smaller for smaller $E_{\rm res}$ values for the reason explained below. As $E_{\rm res}$ decreases, the temperature of the first resonance increases, and that of the second resonance decreases (see Appendix \ref{appendix2_6}).

  \subsubsection{The first resonance}
For 0.007 GeV $\lesssim E_{\rm res} \lesssim 0.04$ GeV, the effective mixing angle for temperatures around the first resonance temperature is given by
\begin{eqnarray}
  \sin^2 2 \tilde{\theta} &=& \frac{\sin^2 2\theta}
      {\sin^2 2\theta +\cos^2 2\theta \left [
          1 +\frac{C_\alpha G_{\rm F}^2 T^4 E^2}{\cos 2 \theta \alpha \delta m^2}
          - \left( \frac{E}{E_{\rm res}} \right)^2 \right]^2} \nonumber \\
      &\approx& \frac{4 \theta^2}{4 \theta^2 +
        \left( \frac{E}{E_{\rm res}} \right)^4 \left [
          \left( \frac{T}{T_{{\rm res},1}} \right)^4 - 1 \right]^2} \nonumber \\
      &\approx& \frac{4 \theta^2}{4 \theta^2 +
        \left( \frac{E}{E_{\rm res}} \right)^4 \left [
          4 \Delta \ln T_1 +6 \left( \Delta \ln T_1 \right)^2 \right]^2},
\end{eqnarray}
where
we took $\Delta \ln T_1 =(T -T_{{\rm res},1}) /T_{{\rm res},1} \ll 1$, and assumed $\theta \ll 1$ and that amplitudes of the second and the third terms in the square brackets in the first line are much larger than unity.
The duration of the resonance, e.g., the full width of temperature at 1/e maximum, is estimated as
\begin{eqnarray}
  \left( \frac{E}{E_{\rm res}} \right)^4 \left [
    4 \Delta \ln T_1 +6 \left( \Delta \ln T_1 \right)^2 \right]^2
    &=&4\theta^2 (e-1) \nonumber \\
    \Longrightarrow \Delta \ln T_1 &\propto& \frac{E_{\rm res}^2}{E^2}.
\end{eqnarray}
Since the sterile neutrino energy at the first resonance has a scaling of
$E =3.15 T_{{\rm res},1} \propto E_{\rm res}^{-1/2}$ [Eq. (\ref{eq_b65})],
the temperature step is given by
$\Delta \ln T_1 \propto E_{\rm res}^{3}$.
When the final abundance is much smaller than the equilibrium abundance, the abundance change at the first resonance roughly scales as
\begin{eqnarray}
  \Gamma_{\nu_{\rm s}}(T_{{\rm res},1}) \Delta t_{{\rm res},1} &\propto&
  \Gamma_{\nu_{\rm s}}(T_{{\rm res},1}) \Delta \ln T_1 H(T_{{\rm res},1})^{-1} \\
  &\propto& T_{{\rm res},1}^5 E_{\rm res}^{3} T_{{\rm res},1}^{-2} \\
  &\propto& E_{\rm res}^{3/2}.
\end{eqnarray}
Therefore, the abundance is smaller for smaller $E_{\rm res}$ values.

\subsubsection{Deactivation of the first resonance}
A discontinuity in contours is seen at a specific energy of $E_{\rm res,cr} \approx 7$ MeV. For $E_{\rm res}>E_{\rm res,cr}$ the codition $T_{{\rm res},1} <T_{\rm eq} =44$ MeV is satisfied, while for $E_{\rm res}<E_{\rm res,cr}$ the codition $T_{{\rm res},1} >T_{\rm eq}$ is satisfied. In the latter case, the sterile neutrino production rate is significantly hindered by small neutrino oscillation phase [Eq. (\ref{eq17})]. The clear discontinuity results from present approximate treatment of Eq. (\ref{eq17}).

\subsubsection{Second resonance}
The temperature step during the second resonance $\Delta \ln T_2$ is constant (see Appendix \ref{appendix2_3}). Using the scaling $T_{{\rm res},2} \propto E_{\rm res}$ [Eq. (\ref{eq_b66})], we obtain a rough scaling of the abundance change for the case that the final abundance is much smaller than the equilibrium abundance, i.e.,
\begin{eqnarray}
  \Gamma_{\nu_{\rm s}}(T_{{\rm res},2}) \Delta t_{{\rm res},2} &\propto&
  \Gamma_{\nu_{\rm s}}(T_{{\rm res},2}) \Delta \ln T_2 H(T_{{\rm res},2})^{-1} \\
  &\propto& E_{\rm res}^{3}.
\end{eqnarray}
Since the first resonance is not effective for $E_{\rm res}<E_{\rm res,cr}$, the abundance change at the second resonance is the final abundance. The abundance is smaller for smaller $E_{\rm res}$ values.

We note that for such a small $E_{\rm res}$ value, the effective mixing angle is much smaller than the bare mixing angle until $E\sim E_{\rm res}$ is realized [Eq. (\ref{eq_add1})]. Therefore, the final abundance is smaller than that of very large $E_{\rm res}$ or the four dimensional model for a fixed $\theta$ value.

\subsection{Decoupling temperature}
Finally, we discuss $\nu_{\rm s}$ decoupling temperature in order to describe time evolution of energy density of the sterile neutrino during BBN. First, we define a parameter
\begin{align}
r_{\rm s}={ \frac{T_{\nu_{\rm s}}}{T_\nu}},
\label{eq18}
\end{align}
where $T_{\nu_{\rm s}}$ and $T_{\nu}$ are temperatures of sterile neutrino and active neutrinos, respectively, for a fixed cosmic time.
This ratio is unity when the sterile neutrino is in equilibrium.  When the sterile neutrino is decoupled, the temperatures can be different, and the ratio is smaller than 1, in general.  After active neutrino decoupling, the two temperatures have the same scaling with a scale factor of the universe.  The ratio is, therefore, kept constant again.

The $r_{\rm s}$ value after the decoupling is given by
\begin{align}
r_{\rm s}=\frac{T_{\nu_{\rm s}}}{T_{\nu}}=\left(\frac{g_{*\mathrm{S}}}{g_{*\mathrm{S,dec}}}\right)^{1/3},
\label{eq19}
\end{align}
where $g_{*\mathrm{S}}$ and $g_{*\mathrm{S,dec}}$ denote the relativistic degrees of freedom, which does not contain the contribution of the sterile neutrino, at the decoupling temperature of active and sterile neutrinos, respectively.  The second equality is derived from the evolution of the active neutrino temperature by taking into account the entropy conservation \cite{Kolb1994}.

This ratio is constant between the initial temperature of BBN calculation, which is taken to be $T_9=100$, and the active neutrino decoupling temperature $T_9 \sim 10$, due to the following reason.  In this temperature interval, numbers of degree of freedom for entropy does not change.  Therefore, both temperatures of the sterile and active neutrinos simply scale as $T\propto 1/a$.  We then use this constant ratio in the BBN calculation.

\begin{figure}
\centering
\includegraphics[width=8cm]{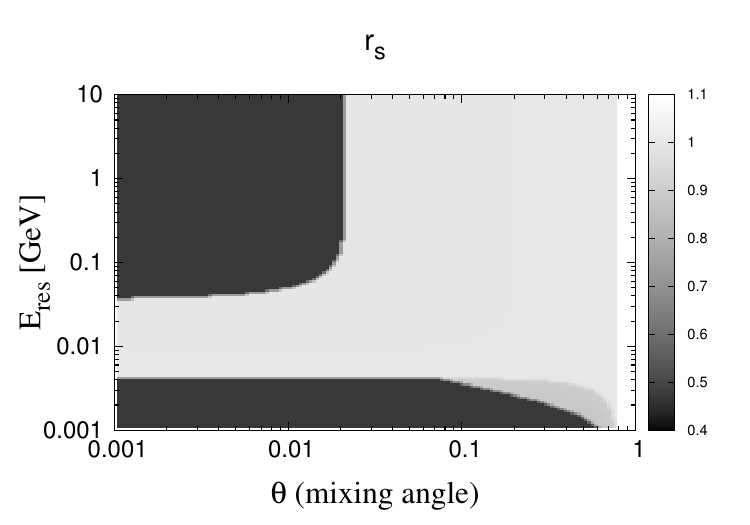}
\caption{(Color online) Color map of $r_\mathrm{s}$ in the parameter plane of $\theta$ and $E_{\rm res}$ [GeV] for $m_{\nu_{\rm s}} =1$ eV.} 
\label{fig5}
\end{figure}

Figure \ref{fig5} shows calculated ratio $r_{\rm s}$ in the parameter plane of $\theta$ and $E_{\rm res}$.  This ratio $r_{\rm s}$ depends on $\theta$ and $E_{\rm res}$ since the decoupling temperature of the sterile neutrino depends on its reaction rate determined by those parameters. The ratio is rapidly increased at the curved boundary. In the light region, the sterile neutrino decouples later than the quark-hadron transition.  The value of $r_{\rm s}$ is therefore close to unity in the region.
  The dark region is corresponding to the small abundance region due to a small reaction rate in Fig. \ref{fig2}.
The final abundance $Y_{\nu_{\rm s}}$ is smaller than the equilibrium abundance at the initial temperature, $Y_{\rm EQ} \sim 0.002$ (see Fig. \ref{fig3}), in that region.  The equilibrium is never realized there.  When the reaction rate does not become larger than the expansion rate, there is no good way of estimation for the temperature of the sterile neutrino.  However, in such a case, the final abundance of the sterile neutrino is always negligibly small, and the temperature is not important.  We then just take the initial temperature $T=100$ GeV as the decoupling temperature for this case.

\section{Results of BBN}
\label{4}
From the rate equation, final abundances and energy density of the sterile neutrino are determined. Then the cosmic expansion rate is modified as follows
\begin{align}
H^2=\frac{8\pi G}{3}(\rho_{\rm standard}+\rho_{\nu_{\rm s}})+\frac{\mathcal{E}}{{ a}^4 ~.}
\label{eq20}
\end{align}
The modified expansion rate changes primordial abundances.  We can then constrain the relevant parameters, $\theta$, $E_{\rm res}$, $m_{\nu_{\rm s}}$ and $\mathcal{E}$, by comparing calculated abundances to the observational data. The primordial elemental abundances depend on the sterile neutrino abundance, $Y_{\nu_{\rm s}}$ shown in Fig. \ref{fig2} because the cosmic expansion rate depends on the energy density of the sterile neutrino.

\subsection{BBN Calculation and Observational Constraints}
 We use updated reaction rates \cite{Descouvemont:2004cw,Coc:2015bhi} in the BBN calculation code  \cite{Kawano1992,Smith:1992yy}. The neutron lifetime is taken from the central value of the Particle Data Group, $880.3 \pm 1.1$~s~\cite{Agashe:2014kda}. The baryon-to-photon ratio is adopted from the value $\eta =(6.037 \pm 0.077) \times 10^{-10}$ corresponding to the baryon density in the base $\Lambda$CDM model (Planck+WP) determined from Planck observation of cosmic microwave background, $\Omega_\mathrm{m} h^2 =0.02205 \pm 0.00028$ \cite{Ade:2013zuv}.

The primordial D abundance comes from observations of quasistellar object (QSO) absorption systems and its value is D/H=$(2.53\pm0.04)\times 10^{-5}$. We take $2\sigma$ limit $(2.53 \pm 0.08) \times10^{-5}$ and $4\sigma$ limit $(2.53 \pm 0.16)\times10^{-5}$ in the following analysis.
For $^4 \mathrm{He}$, we adopt $Y_{\rm p} = 0.2551 \pm 0.0022$ which is observed from metal-poor extragalactic HII region \cite{7} and also consider their $2\sigma$ limit $(0.2551 \pm 0.0044)$ and $4\sigma $ limit $(0.2551 \pm 0.0088)$.

\subsection{$\mathcal{E}=0$ and $m_{\nu_{\rm s}}=1\,\mathrm{keV}$}

\begin{figure}
\centering
\includegraphics[width=7.7cm]{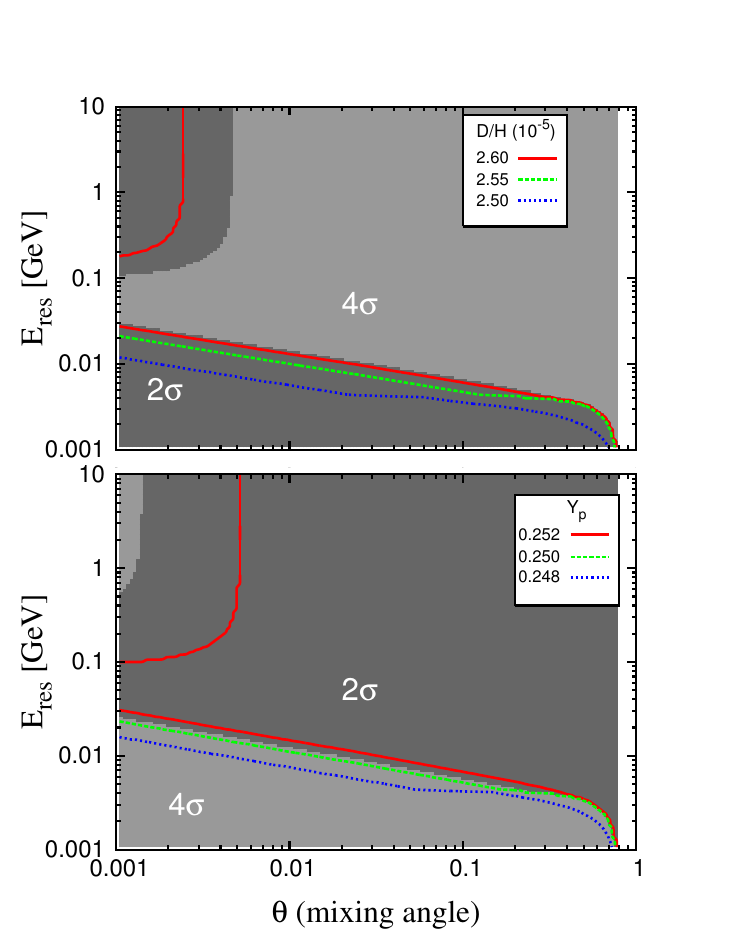}
\caption{(Color online) Contours of deuterium abundance (the top panel) and $^4\mathrm{He}$ mass fraction (the bottom panel) in the parameter plane of $\theta$ and $E_{\rm res}$ [GeV] for the case of $m_{\nu_{\rm s}}=1\,\mathrm{keV}$ and $\mathcal{E}=0$. Parameter ranges are $0.001 \le \theta \le \pi /4$ and $0.001\, \mathrm{GeV} \le E_{\rm res} \le 10\, \mathrm{GeV}$, respectively. Dark and light-shaded regions are 2 $\sigma$ and 4 $\sigma$ allowed regions, respectively.}
\label{fig6}
\end{figure}

Figure \ref{fig6} shows the result of the primordial abundance in the case of $\mathcal{E}=0$ and $m_{\nu_{\rm s}}=1\,\mathrm{keV}$. This $1\,\keV$ scale of the sterile neutrino is one of candidates for dark matter. Since $\mathcal{E}$ is equal to zero, only the energy density of the sterile neutrino affects the cosmic expansion rate. Since the mass of the sterile neutrino is 1\,keV, it is relativistic during BBN epoch. For deuterium abundance, all parameter regions adopted here are allowed by the 4 $\sigma$ abundance limit. We find a parameter region in which the calculated $^4$He abundances satisfy the observational 2$\sigma$ constraints although most of this region does not satisfy the $2\sigma$ limit of D abundance.  This allowed $2\sigma$ region is not seen in the standard BBN result (see Fig. \ref{fig1} at $\mathcal{E}/a_{\rm i}^4 =0$).  The existence of the sterile neutrino energy density, however, increases the cosmic expansion rate, and as a result, abundances of D and $^4$He are increased. All parameter regions in Fig. \ref{fig6} are allowed by the 4$\sigma$ limit.

The shapes of contours can be interpreted as follows. The energy density scales as
\begin{eqnarray}
  \rho_{\nu_{\rm s}} &=& n_{\nu_{\rm s}} \langle E_{\nu_{\rm s}} \rangle \propto Y_{\nu_{\rm s}} T_{\nu_{\rm s}} \nonumber\\
  &\propto& Y_{\nu_{\rm s}} r_{\rm s}.
\label{eq21}
\end{eqnarray}
The number abundance of sterile neutrino is proportional to $Y_{\nu_{\rm s}}$ shown in Fig. \ref{fig2}.\footnote{We note that the values of $Y_{\nu_{\rm s}}$ as well as $r_{\rm s}$ depends on $\delta m^2$. Therefore, contour shapes in Figs.\ref{fig2} and \ref{fig6} are different. The mass assumed for Fig.\,\ref{fig6} is larger than that of Fig. \ref{fig2}. Therefore, the value of $T_{\rm eq}$ is larger. The neutrino oscillation then becomes effective earlier (see Sec. \ref{sec2c} and Appendix \ref{appendix1}).} The abundance is then high in the large $\theta$ region, and there is a narrow peak at $E_{\rm res} = {\mathcal O}(0.01)$ GeV.  Since large values of $r_{\rm s}$ are realized with large $Y_{\nu_{\rm s}}$ values, the factor of $r_{\rm s}$ amplifies the effect of $Y_{\nu_{\rm s}}$.  This dependence is appearing again in Fig. \ref{fig6}.  Since the energy density of sterile neutrino becomes larger for the larger $\theta$ values and the critical resonance energy of $E_{\rm res} = {\mathcal O}(0.01)$ GeV, the abundances of D and $^4$He are also high in that region of Fig. \ref{fig6}.
Also, the energy density is proportional to the sterile neutrino temperature or $r_{\rm s}$ shown in Fig. \ref{fig5}. There is a rapid change of the $r_{\rm s}$ value related to whether the sterile neutrino is decoupled early or not. The value is low at the left bottom and the left top in the parameter space.



\subsection{Constraint on $\mathcal{E}$ }

\begin{figure} [h]
\includegraphics[width=7.5cm]{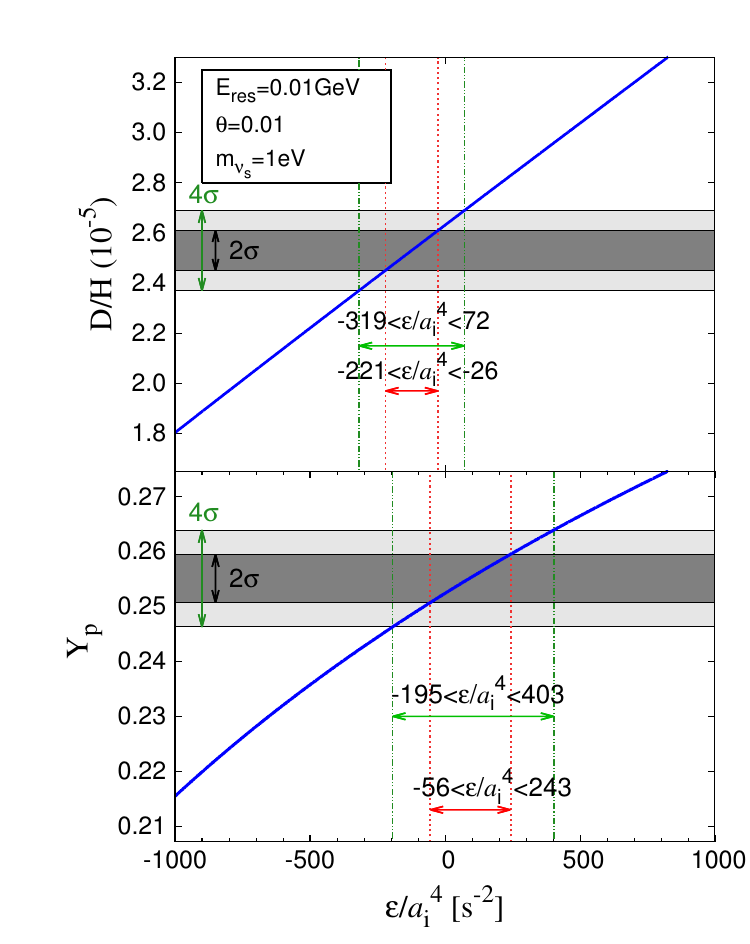}
\caption{Same as Fig.\,\ref{fig1}. But here we include the sterile neutrino which has $m_{\nu_{\rm s}}=1~\mathrm{eV}$, $\theta=0.01$ and $E_{\rm res}=0.01 \mathrm{GeV}$. The value of $\mathcal{E}$ is constrained as $-56<\mathcal{E}/a_{\rm i}^4~({\rm s}^{-2})<-26$ and $-195<\mathcal{E}/a_{\rm i}^4~({\rm s}^{-2})<72$ by the 2$\sigma$ and 4$\sigma$ limits, respectively.}
\label{fig7}
\end{figure}

First, we consider the case of the smallest number abundance of the sterile neutrino realized in the parameter region. Figure \ref{fig7} shows the calculated abundances of deuterium and $^4$He, and also constraints on $\mathcal{E}$ similar to those in Fig. \ref{fig1}.  The mass of the sterile neutrino $m_{\nu_{\rm s}}$ is assumed to be $1\,\mathrm{eV}$ which was the mass scale discussed in the reactor anomalies. Mixing angle $\theta$ and resonance energy $E_{\rm res}$ are fixed to be 0.01 radian and 0.01\,GeV, respectively. From the rate equation result, these values give the lowest reaction rate of the sterile neutrino, that is, the smallest relic number abundance of the sterile neutrino. Since the sterile neutrino increases the cosmic expansion rate, constrained values of $\mathcal{E}$ are shifted to the left side compared to those of Fig.\,\ref{fig1}. Namely, $120<\mathcal{E}/a_{\rm i}^4~({\rm s}^{-2})<149$ ($2\sigma$) and $-18<\mathcal{E}/a_{\rm i}^4~({\rm s}^{-2})<248$ (4$\sigma$) regions are shifted to $-56<\mathcal{E}/a_{\rm i}^4~({\rm s}^{-2})<-26$ (2$\sigma$) and $-195<\mathcal{E}/a_{\rm i}^4~({\rm s}^{-2})<72$ (4$\sigma$), respectively. Therefore, the $2\sigma$ allowed region in Fig.\,\ref{fig1} is totally replaced, and a part of the parameter region of $72<\mathcal{E}/a_{\rm i}^4~({\rm s}^{-2})<248$ (4$\sigma$) are excluded by the sterile neutrino existent in BBN epoch.

\begin{figure} [h]
\includegraphics[width=7.5cm]{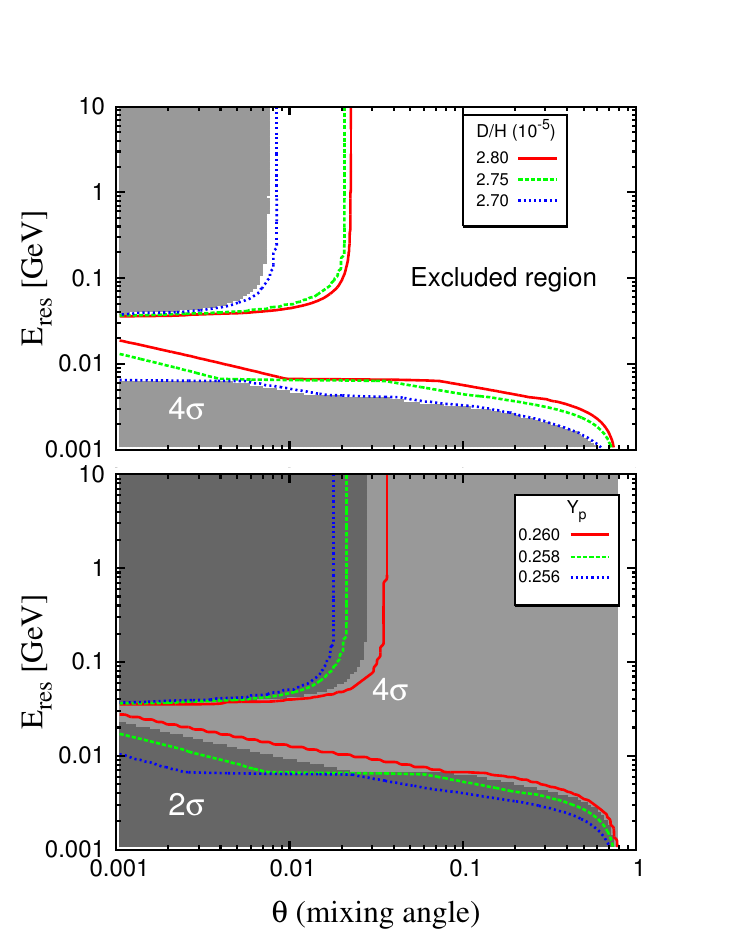}
\caption{(Color online) Same as Fig. \ref{fig6}, but for $\mathcal{E}/a_{\rm i}^4 =248$ s$^{-2}$ and $m_{\nu_{\rm s}}=1\,\mathrm{eV}$.} 
\label{fig8}
\end{figure}

Fig.\,\ref{fig8} shows the contours for the case of $\mathcal{E}/a_{\rm i}^4 =248$ s$^{-2}$ which is the maximum value of $\mathcal{E}$ in the 4 $\sigma$ allowed region in Fig.\,\ref{fig1} and $m_{\nu_{\rm s}}=1\,\mathrm{eV}$. This $\mathcal{E}$ value is excluded by the over-abundance of deuterium when the 1 eV sterile neutrino with $\theta=0.01$ and $E_{\rm{res}}=0.01\,\GeV$ is added.

vi


\subsection{Constraint on the mass}

\begin{figure} [h]
\includegraphics[width=7.5cm]{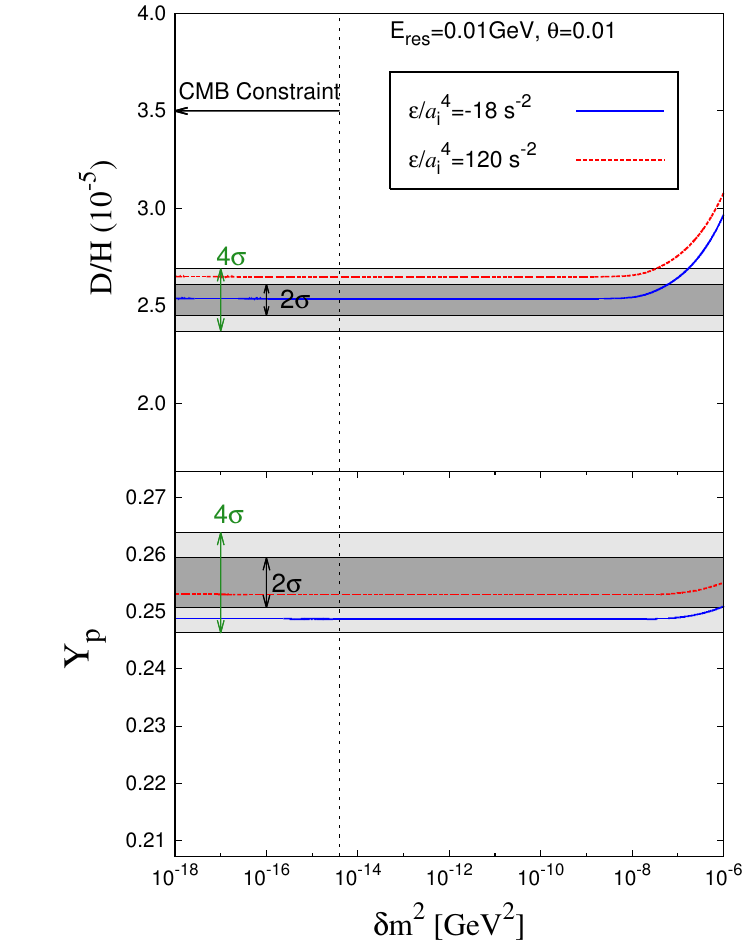}
\caption{(Color online) Primordial abundances of D and $^4$He as a function of $\delta m^2$. The parameters $\theta$ and $E_{\rm res}$ are fixed at 0.01 radian and 0.01 GeV, respectively. The solid blue and dotted red lines denote the abundances for $\mathcal{E}/a_{\rm i}^4 =-18$ and $120$ s$^{-2}$, respectively. The limits from observed abundances are delineated by the horizontal lines: inner lines ($2\sigma$) and  outer lines ($4\sigma$). In the right region from the vertical black-dashed line, the present energy density of the sterile neutrino is larger than the constraint from CMB observation.}
\label{fig9}
\end{figure}

Figure \ref{fig9} shows the primordial abundances as a function of $\delta m^2 \equiv m_{\nu_{\rm s}}^2 - m_{\nu_a}^2$ in the same condition of Fig.\,\ref{fig7}, i.e., $\theta=0.01$ and $E_{\rm res} =0.01$ GeV. The value of $\mathcal{E}$ is fixed by the lowest values in the 2 $\sigma$ and 4$\sigma$ allowed regions for the case without the sterile neutrino. Assuming $m_{\nu_a} \ll 1\,\mathrm{eV}$, we neglected the mass of the active neutrino. Thus, $\delta m^2$ is approximately the same as the squared mass of the sterile neutrino. For $\delta m^2 \lesssim 10^{-9}\,\mathrm{GeV}{^2}$, there is no contribution of the sterile neutrino mass because the sterile neutrino is relativistic during the BBN epoch. If the sterile neutrino is relativistic in the BBN epoch, then the cosmic expansion rate does not depend on the mass of the sterile neutrino but only its number density.

However, if $m_{\nu_{\rm s}}$ is larger, then the sterile neutrino would be non-relativistic and its mass affects the cosmic expansion rate.
As a result, for $\mathcal{E}/a_{\rm i}^4=-18$ s$^{-2}$, $\delta m^2$ is allowed up to $1.7 \times10^{-7}\,\mathrm{GeV^2}$ by the 4 $\sigma$ constraint. On the other hand, there is no allowed region for the $2\sigma$ range.
For $\mathcal{E}/a_{\rm i}^4=120$ s$^{-2}$, $\delta m^2$ is allowed up to $ 3.3 \times 10^{-8}\,\GeV^2$ for the $4\sigma$ range and also there is no allowed region for the $2\sigma$ range.

In addition, if the relic sterile neutrino can be existed on the brane in the present universe, it can be a candidate of dark matter. We can constrain it from the observational data of cosmic microwave background (CMB).
  However, we do not know what happens during BBN and the present time in the extra-dimensional universe.  Perhaps the sterile neutrino may diffuse in the extra-dimensional bulk associated with bulk expansion which is beyond the scope of this paper and not treated in this study.  We should then note that what we derive in this paper is a constraint independently coming from the BBN consideration alone on the physical environment in the short BBN epoch.

  The \textit{Planck} observation gives the following data of cold dark matter density parameter for $\Lambda \rm{CDM}$ model with \textit{Planck} temperature power spectrum data alone \cite{Ade:2013zuv}:
\begin{equation}
\Omega_c h^2 = 0.1196 \pm 0.0031.
\label{eq_ad1}
\end{equation}
This corresponds to the energy density of the cold dark matter $\rho_c$
\begin{equation}
\rho_c=(0.1261 \pm 0.0033) \times 10^{-5}\, \GeV\, \rm{cm}^{-3}.
\label{eq_ad2}
\end{equation}
The present energy density of the sterile neutrino $\rho_{\nu_{{\rm s}0}}$ cannot be larger than the observed energy density of dark matter.  Therefore, if the relic sterile neutrino totally remains on our brane until now, we have a constraint of $\rho_{\nu_{{\rm s}0}} \le \rho_c$ that leads to
\begin{eqnarray}
r_{\rm s} Y_{{\nu_{\rm s}0}} &\le& (8.257 \pm 0.22) \times 10^2 \ (\mathrm{for}~\langle E_{\nu_{{\rm s}0}} \rangle > m_{\nu_{\rm s}})~, \\
  m_{\nu_{\rm s}} Y_{{\nu_{\rm s}0}} &\le& (4.362 \pm 0.11) \times 10^{-10}~\rm GeV \nonumber \\
  && \hspace{20ex} (\mathrm{for}~\langle E_{\nu_{{\rm s}0}} \rangle < m_{\nu_{\rm s}}),~~~
\label{eq_ad5}
\end{eqnarray}
where $\left\langle E_{\nu_{{\rm s}0}} \right\rangle = 3.151 (4/11)^{1/3} T_{\gamma 0}$ is the average present temperature of the sterile neutrino when it is massless, with $T_{\gamma 0} =2.7255$ K the present CMB temperature \cite{Fixsen:2009}. The first and second lines correspond to constraints on the relativistic and nonrelativistic sterile neutrinos, respectively.  In our calculation, the maximum value of $r_{\rm s}$ and $Y_{\nu_{\rm s}}$ are $\sim 1$ and $\sim 0.02$, respectively. Thus, all parameter space for the relativistic case are allowed by the CMB data. For the non-relativistic case, since the energy density of the sterile neutrino is proportional to the mass, the allowed region from the CMB data becomes narrow with increasing sterile neutrino mass.  We choose $m_{\nu_{\rm s}} = 100\,\keV$, i.e., the mass scale with which the sterile neutrino becomes non-relativistic at the typical BBN temperature of $T_9=1$.  For this mass value, the region of $E_{\rm{res}} \lesssim \mathcal{O}(0.001) \GeV$ is only allowed.  As seen in Fig.\,\ref{fig9}, the constraint from the CMB data sets the upper limit on $m_{\nu_{\rm s}}$ at $\delta m^2 =\mathcal{O}(10^{-15})$ GeV$^2$.

\begin{figure} [h]
\includegraphics[width=7.5cm]{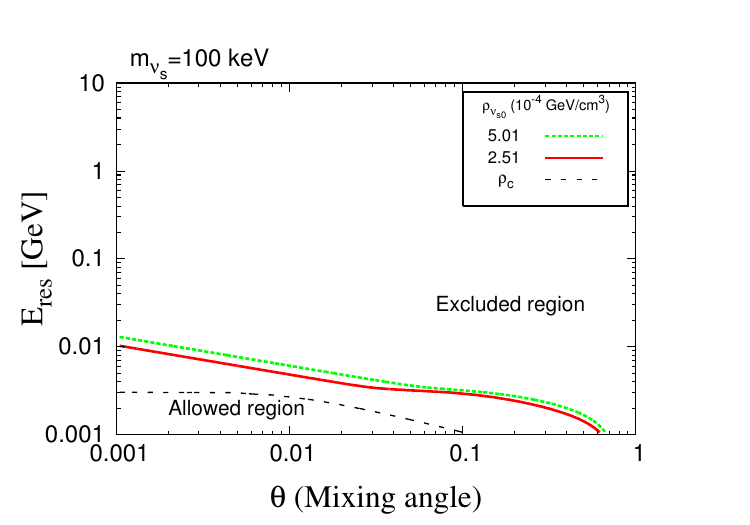}
\caption{(Color online) Contours of the energy density of relic sterile neutrino in the parameter plane of $\theta$ and $E_{\rm{res}}$.  The mass of the sterile neutrino is 100\,\rm{keV}.  The black dashed line corresponds to the energy density of cold dark matter at present deduced from the \textit{Planck} observational data.  The right upper region from this line is excluded. The red-solid and green-dashed lines are also shown to be excluded by the CMB observational data.}
\label{fig10}
\end{figure}

  Figure \ref{fig10} shows the contours for the present energy density of the relic sterile neutrino deduced from calculated results of $Y_{\nu_{\rm s}}$ in the parameter plane of $\theta$ and $E_{\rm{res}}$ for $m_{\nu_{\rm s}} =100$ keV. The black dashed line corresponds to the present energy density of cold dark matter [Eq. (\ref{eq_ad2})].  The right upper region from this line is excluded.

\begin{figure} [h]
\includegraphics[width=7.5cm]{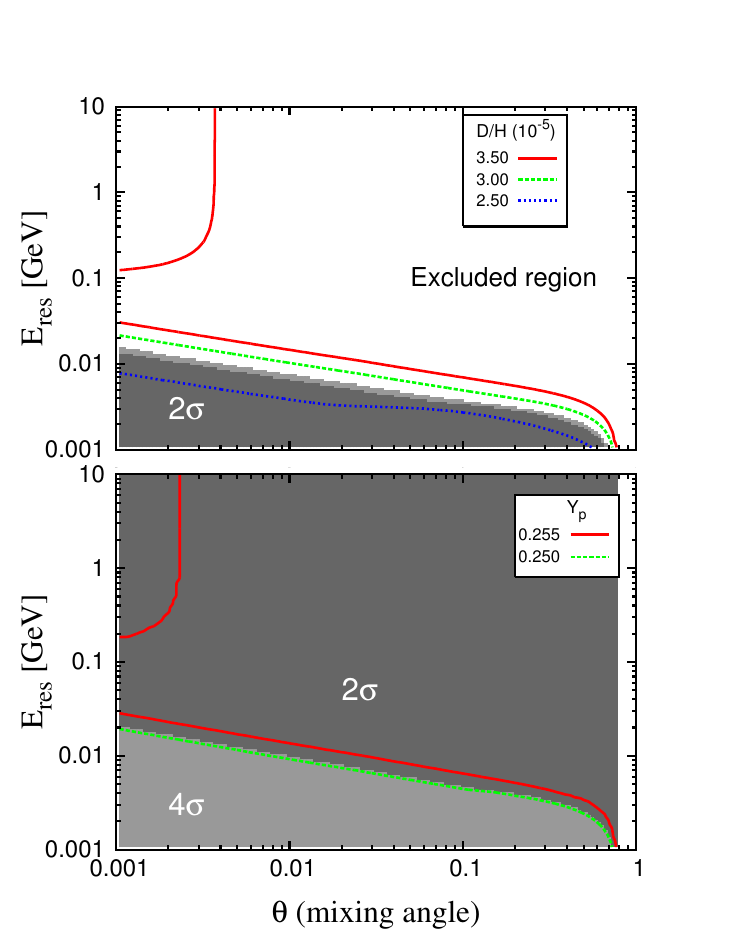}
\caption{(Color online) Same as Fig. \ref{fig6}, but for $m_{\nu_{\rm s}}=1~\mathrm{MeV}$ and $\mathcal{E}/a_{\rm i}^4=-20$ s$^{-2}$.}
\label{fig11}
\end{figure}

Figure \ref{fig11} shows the same contours of light element abundances as in Fig. \ref{fig6} for the case of 1 MeV sterile neutrino and $\mathcal{E}/a_{\rm i}^4=-20$ s$^{-2}$, for example.  This value of $\mathcal{E}/a_{\rm i}^4=-20$ s$^{-2}$ is near the lowest allowed value in Fig.\,\ref{fig1} for the 4\,$\sigma$ limit. In this case, there is no parameter region that satisfies both of the $2\sigma$ limits on D/H and $Y_{\rm p}$.  The $4\sigma$ allowed region is located in the left-bottom region.


In the present model of a sterile neutrino, the effective mixing angle depends on the energy by Eq.\,(\ref{eq15}). (See also Figs.\,\ref{fig_b2} and \ref{fig_b3} in Appendix B.) Results of neutrino experiments, therefore, do not always exclude the parameter region for large values of $\theta$.  For example, if we assume the resonance energy $E_{\rm res}=400$ MeV (corresponding to Figure 4 in Ref. \cite{Pas:2005rb}), for the energy region of the IceCube measurement \cite{TheIceCube:2016oqi}, i.e., 320 GeV $\sim$ 20 TeV, the effective mixing angle becomes negligibly small. The IceCube data is, therefore, consistent with this model, independently of the mixing angle $\theta$. The experimental verification of the mixing of a sterile neutrino, which propagates to the bulk space, then requires measurements of the effective mixing angle for various neutrino energies.

\subsection{Dependence of primordial abundances on $\delta m^2$ and $\theta$}
\begin{figure} [h]
\includegraphics[width=7.5cm]{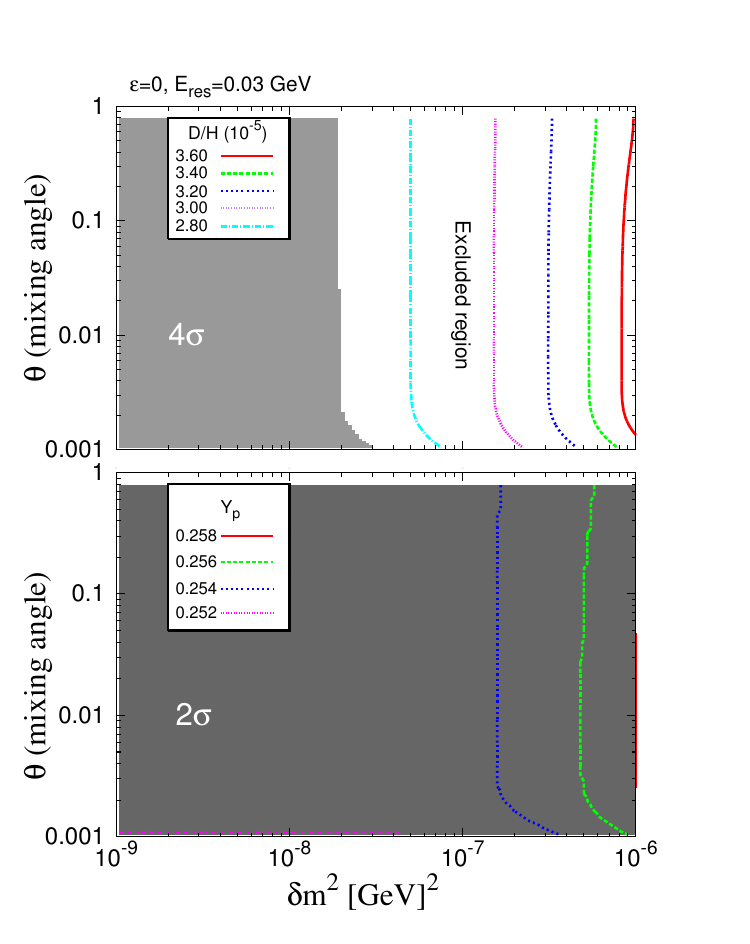}
\caption{(Color online) Contours of deuterium abundance (the top panel) and $^4\mathrm{He}$ abundance (the bottom panel) in the parameter plane of $\theta$ and $\delta m^2 \,[\mathrm{GeV}^2]$ for the case of $\mathcal{E}=0$ and $E_{\rm res} = 0.03 \, \mathrm{GeV}$. Parameter ranges are $0.001 \leq \theta \leq \pi/4$ and $1.0 \times 10^{-9} \, \mathrm{GeV}^2 \leq \delta m^2 \leq 1 \times 10^{-6}\,\mathrm{GeV}^2$. Dark- and light-shaded regions are the $2\sigma$ and $4\sigma$ allowed regions, respectively. The white region is excluded by the BBN constraint.}
\label{fig12}
\end{figure}

Figure \ref{fig12} shows the primordial abundances as a function of $\delta m^2$ and the mixing angle $\theta$. The values of $\mathcal{E}$ and $E_{\rm res}$ are fixed, respectively, at 0 and 0.03 GeV. If $\delta m^2$ is larger than $10^{-8}\,\mathrm{GeV}^2$, the primordial abundances are increased because the cosmic expansion rate depends on the mass of the sterile neutrino by Eq.\, (\ref{eq7}). In the high mass region of the figure, deuterium and $^4\mathrm{He}$ abundances are high.  If the mixing angle is increased, the reaction rate of the sterile neutrino is also increased.  As a result, higher number- and energy-densities of the sterile neutrino are obtained. Therefore, final abundances of light elements become higher by increasing the mixing angle $\theta$ similarly to the trend in Fig. \ref{fig9}. If the $\delta m^2$ value is higher than $(2-3)\times 10^{-8}\, \mathrm{GeV}^2$, the region is excluded by over-production of the deuterium. Similarly to the case of Fig.\,\ref{fig9}, only the relativistic mass region is allowed by the CMB observational data in this case. 
\\

\subsection{Dependence of primordial abundances on $\delta m^2$ and $E_{\rm res}$}
\begin{figure} [h]
\includegraphics[width=7.5cm]{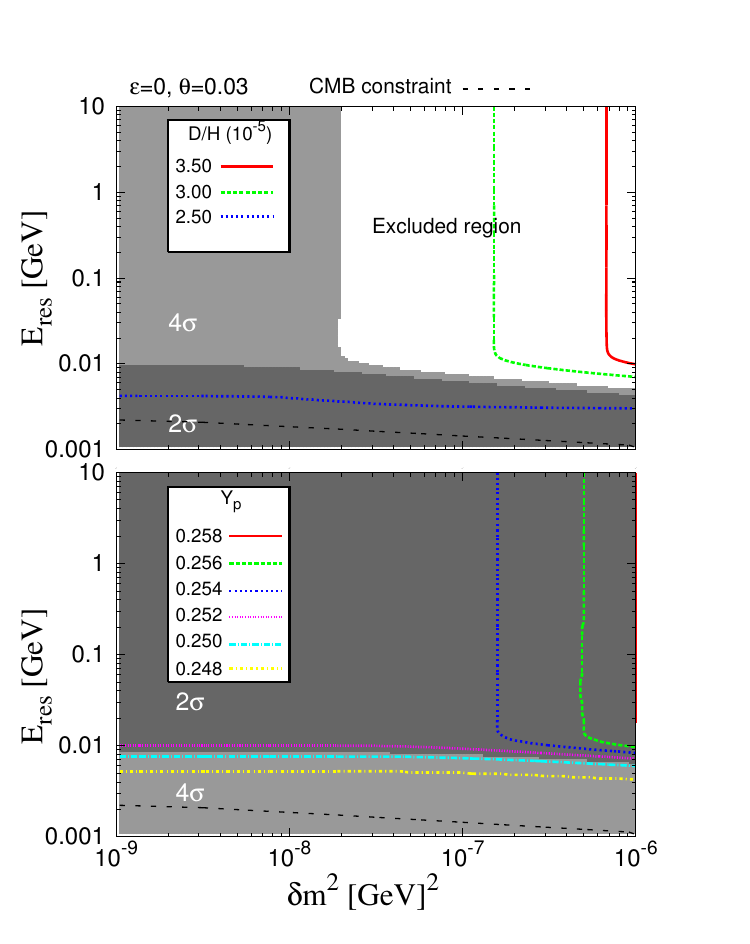}
\caption{(Color online) Contours of deuterium abundance (the top panel) and $^4\mathrm{He}$ abundance (the bottom panel) in the parameter plane of $E_{\rm res}$ and $\delta m^2 \,[\mathrm{GeV}^2]$ for the case of $\mathcal{E}=0$ and $\theta = 0.03$. Parameter ranges are $0.001\,\mathrm{GeV} \leq \mathrm{E_{res}} \leq 10\,\mathrm{GeV}$ and $1.0 \times 10^{-9} \, \mathrm{GeV}^2 \leq \delta m^2 \leq 1.0 \times 10^{-6}\,\mathrm{GeV}^2$. Dark- and light-shaded regions are the $2\sigma$ and $4\sigma$ allowed regions, respectively. White region is excluded by the BBN constraint. The black dashed line corresponds to the CMB constraint on the present energy density of dark matter.}
\label{fig13}
\end{figure}

Figure \ref{fig13} shows the primordial abundances as a function of $\delta m^2$ and $E_{\rm res}$. The value of $\mathcal{E}$ and $\theta$ are fixed, respectively, at 0 and $0.03\,\GeV$. In the figure, the dependence on $\delta m^2$ is similar to Fig. \ref{fig12} by the same reason. The curved shape at the specific $E_{\rm res}$ value appears by the following reason:  The number abundance of the sterile neutrino is the highest when the resonance energy is equal to the energy of the sterile neutrino during its decoupling epoch. Then the higher number abundance of sterile neutrino around $E_{\rm res} \simeq T_{\rm eq} \sim 0.04\,\mathrm{GeV}$ makes the larger energy density.  It affects the cosmic expansion rate more and results in higher primordial abundances. The curved shape is then similar to the pattern in Fig. \ref{fig2}. The region of $\delta m^2 \gtrsim 2\times 10^{-8}\,\mathrm{GeV}^2$ and $E_{\rm res} \gtrsim 0.01$ GeV is excluded by the over-production of deuterium, also in this case.
We find a parameter region for the $2\sigma$ allowed region at $E_{\rm res}\sim 0.01$ GeV and $\delta m^2 \lesssim 10^{-8}$ GeV$^2$.

\section{Conclusion}
\label{5}
In this work, we study effects of a sterile neutrino which can propagate in the bulk and brane in the five dimensional universe on BBN.  In the present model, the cosmic expansion rate is modified by the energy density of the sterile neutrino and the existence of the fifth dimension itself.  We then deduce parameter regions relevant to the multi-dimensional sterile neutrino by using results of the BBN calculation. The five-dimensional effect is described by one parameter $\mathcal{E}$, and the energy density of the sterile neutrino depends on three parameters, i.e., $E_{\rm res}$, $\theta$ and $m_{\nu_{\rm s}}$.  This model therefore has four physical parameters. Two of them are integration constants: (1) $\mathcal{E}$ comes from the integration of five-dimensional Einstein equation and (2) $\epsilon_{\rm s}$  describes the shortcut, i.e., the difference of geodesics in the bulk and on the brane in five dimensional cosmology. The latter is reflected in the sterile neutrino resonance energy $E_{\rm res}$ in Eq. (\ref{eq16}). The other two parameters are mixing angle and mass scale of the sterile neutrino. These four parameters modify the cosmic expansion rate and the energy density in the BBN epoch. Taking into account the modified cosmic expansion rate, we investigated how primordial abundances are changed and constrained the parameters using the observational abundance data.

First, the parameter $\mathcal{E}$ manifests itself in the cosmic expansion rate and influences the primordial abundances. When we do not consider the sterile neutrino, the paremeter $\mathcal{E}$ is constrained: $120< \mathcal{E}/a_{\rm i}^4~({\rm s}^{-2})<149$ and $-18< \mathcal{E}/a_{\rm i}^4~({\rm s}^{-2})<248$ from the observational 2 $\sigma$ and 4 $\sigma$ limits, respectively, on abundances (Fig. \ref{fig1}).

Second, we took into account the effect of the energy density of the sterile neutrino which can propagate in the bulk space.  The relic abundance and the temperature of the sterile neutrino are calculated by solving the rate equation. The energy density of the sterile neutrino depends on not only the mass but also the number density and the temperature.  Since the mixing angle and the resonance energy are related to the reaction rate of the sterile neutrino, the two parameters determine the relic abundance of the sterile neutrino.   The parameters  are then constrained through the comparison of the BBN calculation results and observed elemental abundances.  The final abundance of the sterile neutrino is increased when the sterile neutrino has a large reaction rate. 

We found that the relic abundance is large for large values of $\theta$ and a characteristic resonance energy $E_{\rm res} \sim 0.04$ GeV (Fig. \ref{fig2}).  This value of resonance energy corresponds to the temperature at which the average scattering time scale equals to the overlap time scale of wave packets for active neutrinos.  The ratio of the temperatures of the sterile and active neutrinos, $r_{\rm S}$, after the decoupling of the active neutrino depends on the decoupling temperature of the sterile neutrino.  The decoupling temperature is determined by the parameters $\theta$ and $E_{\rm res}$.  It is found that the ratio is significantly changed depending on whether the decoupling occurs before or after the quark hadron transition (Fig. \ref{fig5}).

When the sterile neutrino is taken into account, the cosmic expansion rate is increased and high $\mathcal{E}$ values are excluded.  For example, we observed that the constraints on $\mathcal{E}$ in Fig. \ref{fig1} are shifted to $-56< \mathcal{E}/a_{\rm i}^4~({\rm s}^{-2})<-26$ ($2\sigma$) and $-195< \mathcal{E}/a_{\rm i}^4~({\rm s}^{-2})<72$ ($4\sigma$) in the case of $m_{\nu_{\rm s}}=1$ eV, $\theta=0.01$ and $E_{\rm res}=0.01$ GeV (Fig. \ref{fig7}).

When the sterile neutrino is relativistic during BBN, the energy density of the sterile neutrino is determined by the relic abundance $Y_{\nu_{\rm s}}$ and the temperature ratio $r_{\rm S}$.  The energy density is larger for a sterile neutrino which decouples later since its abundance and the temperature ratio are larger.  We then derived a constraint on the parameters for the case of $\mathcal{E} =0$ (Fig. \ref{fig6}).
If the mass of the sterile neutrino is larger than $\sim$1 MeV, then it becomes non-relativistic in the BBN epoch.  So the energy density of the sterile neutrino is proportional to its mass. It gives a large energy density and it is constrained strongly.  For the case of $m_{\nu_{\rm s}} \geq 1\,\mathrm{MeV}$, there are no allowed parameter region consistent with the 2 $\sigma$ limit from observational data (Fig. \ref{fig11}).

We showed a result of a parameter search in the plane of ($\theta$, $m_{\nu_{\rm s}}$) for a fixed $E_{\rm res}$ value.  We found that the region of $m_{\nu_{\rm s}} \gtrsim \mathcal{O}(10^{-4})\,\GeV$ is excluded within all mixing angle parameter space searched in this study when $E_{\rm res} = 0.03\,\GeV$. On the other hand, all mixing angle parameter space are allowed when $m_{\nu_{\rm s}} \lesssim \mathcal{O}(10^{-4})\,\GeV$ within $4\,\sigma$ range. This is because the heavier mass leads to a larger energy density, and the larger mixing angle leads to a later decoupling and a larger number density.  In both cases, the energy density is larger, and that parameter region is constrained (Fig. \ref{fig12}).    We also showed a result of a parameter search in the plane of ($E_{\rm res}$, $m_{\nu_{\rm s}}$) for a fixed $\theta$ value.  We then checked trends of large effects for larger mass and the characteristic resonance energy (Fig. \ref{fig13}).

\begin{acknowledgments}
This work is supported by
the National Research Foundation of Korea \ (Grant No. NRF-2014R1A2A2A05003548 and NRF-2015K2A9A1A06046598).
MK is supported by JSPS Postdoctoral Fellowship for Research Abroad.
\end{acknowledgments}

\appendix
\section{Flavor-change probability $P_{\rm{as}}$}\label{appendix1}

When the matter effect \cite{Notzold:1988,Barbieri:1989ti,Dolgov:2001} is taken into account in the current five-dimensional model \cite{Pas:2005rb}, the effective mixing angle becomes
\begin{eqnarray}
  \sin^2 2 \tilde{\theta} &=& \frac{\sin^2 2\theta}
      {\sin^2 2\theta +\cos^2 2\theta \left [
          1 +\frac{C_\alpha G_{\rm F}^2 T^4 E^2}{\cos 2 \theta \alpha \delta m^2}
          - \left( \frac{E}{E_{\rm res}} \right)^2 \right]^2} \nonumber \\
      &=& \frac{\sin^2 2\theta}{Q_\alpha^2(\theta, \delta m^2, E_{\rm res}; T, E)},
      \label{eq_ap1}
\end{eqnarray}
where $\alpha$ is the fine structure constant, and
$C_e =1.22$ (for $\nu_e$) and $C_{\mu,\tau} =0.34$ (for $\nu_\mu$ and $\nu_\tau$) are flavor ($\alpha$) dependent constants.
In the second equality, we defined a modification factor for the mixing angle by the matter and the extra-dimension effects, i.e.,
\begin{widetext}
\begin{equation}
  Q_\alpha(\theta, \delta m^2, E_{\rm res}; T, E) =
  \sqrt{\mathstrut \sin^2 2\theta +\cos^2 2\theta \left [
    1 +\frac{C_\alpha G_{\rm F}^2 T^4 E^2}{\cos 2 \theta \alpha \delta m^2}
    - \left( \frac{E}{E_{\rm res}} \right)^2 \right]^2
  }.
  \label{eq_ap2}
\end{equation}
\end{widetext}

The probability of the flavor change of $\nu_{\rm a} \leftrightarrow \nu_{\rm s}$ after propagation of time $t$ \cite{Pas:2005rb} taking into account the evolution of the wave packet \cite{Giunti:1991,Fukugita:2010} is given
by
\begin{widetext}
\begin{equation}
  P_{\rm as} = \frac{1}{2} \sin^2 2 \tilde{\theta}
  \left\{ 1 -\cos \left( \frac{\delta m^2_{\rm mat} t}{2 E} \right)
  \exp \left[
    - \left( \frac{t}{L^{\rm coh}_{\rm mat}} \right)^2
    -\left( 1+\kappa \right) \frac{ \left( \delta m^2_{\rm mat} \right)^2}{32 \sigma_p^2 p^2 }
    \right] \right\},
  \label{eq_ap3}
\end{equation}
\end{widetext}
where
$\delta m^2_{\rm mat}$ is given by
\begin{eqnarray}
\delta m^2_{\rm{mat}} = \delta m^2 Q_\alpha(\theta, \delta m^2, E_{\rm res}; T, E).
  \label{eq_ap4}
\end{eqnarray}
The coherent length $L^{\rm coh}_{\rm mat}$ is defined \cite{Giunti:1992} as
\begin{widetext}
\begin{equation}
  L^{\rm coh}_{\rm mat} = L^{\rm coh}_{\rm vac}
  \left|\frac{\delta m^2_{\rm mat}}{\delta m^2  + \cos 2\theta \left[
      \frac{C_\alpha G_{\rm F}^2 T^4 E^2}{\alpha}
      - \cos 2\theta \delta m^2 \left( \frac{E}{E_{\rm res}} \right)^2
      \right]} \right|,
  \label{eq_ap5}
\end{equation}
\end{widetext}
where the coherent length in vacuum is given by
\begin{equation}
  L^{\rm coh}_{\rm vac} = 2 \sqrt{\mathstrut 2} \sigma_x \frac{2 p^2}{\delta m^2}.
  \label{eq_ap6}
\end{equation}
The quantity $\kappa$ is given by
\begin{equation}
  \kappa \approx \frac{p_1^2 - p_2^2}{\delta m^2_{\rm mat}},
  \label{eq_ap7}
\end{equation}
with $p_1$ and $p_2$ average momenta of mass eigenstates 1 and 2, respectively,
and
$\sigma_x$ and $\sigma_p$ are widths of position and momentum, respectively.  There is a relation of $\sigma_x \sigma_p =1/2$.

The first term in Eq. (\ref{eq_ap3}) corresponds to the contribution of squared terms of mass eigenstates 1 and 2, while the second oscillation term with damping corresponds to the interference term of states 1 and 2.

Taking $p \sim \sigma_p = 1/(2\sigma_x) \sim T$ \cite{Barbieri:1989ti}, because of $\kappa \leq T^2 /\delta m^2_{\rm mat}$, the amplitude of second term in the exponential in Eq. (\ref{eq_ap3}) is
\begin{equation}
  \left( 1+\kappa \right) \frac{ \left( \delta m_{\rm mat}^2 \right)^2}{32 \sigma_p^2 p^2 }
  \leq
  \frac{\delta m^2_{\rm mat}}{T^2}.
  \label{eq_ap8}
\end{equation}
We assume that this factor is always much less than unity, and can be neglected in Eq. (\ref{eq_ap3}).

We note that the coherence length $L^{\rm coh}_{\rm mat}$ for $\theta \ll 1$ is roughly given by that of the vacuum oscillation excepting a region where the sum of the second and third terms in the square brackets in Eq. (\ref{eq_ap1}) is comparable to unity.  When the correction by the matter potential plus the extra-dimensional term is dominant, $L^{\rm coh}_{\rm mat} =L^{\rm coh}_{\rm vac} /\cos 2\theta$.  On the other hand, when the correction is negligible, $L^{\rm coh}_{\rm mat} =L^{\rm coh}_{\rm vac}$ is realized.  We then approximate the coherent length by that of the vacuum oscillation.  Using approximations above, the flavor change probability is given by
\begin{widetext}
\begin{equation}
  P_{\rm as} = \frac{1}{2} \sin^2 2 \tilde{\theta}
  \left\{ 1 -\cos \left( \frac{\delta m^2_{\rm mat} t}{2 E} \right)
  \exp \left[
    - \left( \frac{t}{L^{\rm coh}_{\rm vac}} \right)^2
    \right] \right\}.
  \label{eq_ap9}
\end{equation}
\end{widetext}

The production rate of the sterile neutrino is given by
\begin{equation}
  \Gamma_{\nu_{\rm s}}=\Gamma_{\rm w} P_{\rm as}.
  \label{eq_ap10}
\end{equation}

This production rate is evaluated with the mean life of active neutrino against destruction via the weak interaction \cite{Barbieri:1989ti}.  The mean life is given \cite{Barbieri:1989ti} by the average scattering time scale,
\begin{equation}
  t_{\rm sc} \simeq \frac{1}{G_{\rm F}^2 T^5}.
  \label{eq_ap11}
\end{equation}
This time scale is shorter than the cosmic expansion time scale before the active neutrino decoupling.

The coherent length is
\begin{equation}
  L^{\rm coh}_{\rm vac} = 2 \sqrt{\mathstrut 2} \sigma_x \frac{2 p^2}{\delta m^2} \sim \frac{T}{\delta m^2}.
  \label{eq_ap12}
\end{equation}
This is equivalent to the overlap time scale of neutrino wave packets $t^{\rm coh} =L^{\rm coh}_{\rm vac}$.

\subsection{Matter effect}
First, we consider the neutrino oscillation in the case without the extra-dimensional correction.
The ratio of the two different time scales are given by
\begin{eqnarray}
  \frac{t^{\rm coh}}{t_{\rm sc}} &\simeq& \frac{T /\delta m^2}{1/(G_{\rm F}^2 T^5)} \nonumber \\
  &=& \frac{G_{\rm F}^2 T^6}{\delta m^2}.
  \label{eq_ap13}
\end{eqnarray}
Then, the time scales are comparable at the temperature of
\begin{eqnarray}
  T_{\rm eq} &=&\left( \frac{\delta m^2}{G_{\rm F}^2} \right)^{1/6} \nonumber \\
  &=& 44~{\rm MeV} \left(\frac{\delta m^2}{1~{\rm eV}^2} \right)^{1/6}.
  \label{eq_ap14}
\end{eqnarray}

Then we obtain $t^{\rm coh} \ge t_{\rm sc}$ for $T \ge T_{\rm eq}$ and $t_{\rm sc} \ge t^{\rm coh}$ for $T \le T_{\rm eq}$.  Therefore, the coherence survives for $T \ge T_{\rm eq}$, while it is lost for $T \le T_{\rm eq}$.
We note that at $T_{\rm eq}$, the matter term in $\delta m^2_{\rm mat}$ becomes
\begin{eqnarray}
  \frac{C_\alpha G_{\rm F}^2 T_{\rm eq}^4 {E_{\rm eq}^2}}{\cos 2 \theta \alpha \delta m^2}
  \sim \frac{1}{\alpha}.
  \label{eq_ap15}
\end{eqnarray}
This temperature thus roughly corresponds to the epoch when the matter effect becomes unimportant.

The flavor change probability then scales as
\begin{equation}
  P_{\rm as} \approx \left\{
  \begin{array}{ll}
    \frac{1}{2} \sin^2 2 \tilde{\theta}
    \left\{ 1 -\cos \left( \frac{\delta m^2_{\rm mat} t_{\rm sc}}{2 E} \right) \right\} \\
    ~~~~~=\sin^2 2 \tilde{\theta} \sin^2 \left( \frac{\delta m^2_{\rm mat} t_{\rm sc}}{4 E} \right)
    & ({\rm for}~ T \geq T_{\rm eq}) \\
    \frac{1}{2} \sin^2 2 \tilde{\theta}
    & ({\rm for}~ T \leq T_{\rm eq}).
  \end{array}
  \right.
  \label{eq_ap16}
\end{equation}
We thus find that after the typical temperature $T_{\rm eq}$ the flavor change probability does not oscillate since the coherence is lost during the propagation.

In the early epoch of $T \geq T_{\rm eq}$, the oscillation phase reduces to
\begin{eqnarray}
  \frac{\delta m^2_{\rm mat}  t_{\rm sc}}{4E} &\simeq&
  \frac{ \delta m^2}{4E}
  \left( \frac{C_\alpha G_{\rm F}^2 T^4 E^2}{\alpha \delta m^2} \right)
  \left( \frac{1}{G_{\rm F}^2 T^5} \right) \nonumber \\
  &\simeq & \frac{1}{\alpha}.
  \label{eq_ap17}
\end{eqnarray}
The oscillation is, therefore, maximally operative.  We can then take the time average of the probability.  As a result, the flavor change probability for any temperature is given by
\begin{equation}
  P_{\rm as} \approx \frac{1}{2} \sin^2 2 \tilde{\theta}.
  \label{eq_ap18}
\end{equation}

\subsection{Extra-dimensional effect}
Second, we consider the effect of the extra-dimension.  If the term $(E/E_{\rm res})^2$ in Eq.\,(\ref{eq_ap1}) effectively increases the effective mixing angle, the flavor change probability can increase.  The sterile neutrino production rate $\Gamma_{\nu_{\rm s}}$ is then increased.  However, when the factor $Q_\alpha$ is significantly decreased by the extra-dimensional term, the term $\delta m^2_{\rm mat}$ becomes small.  Therefore, the approximation of the maximal oscillation can be broken.  In an extreme case when the factor $Q_\alpha$ is very small, the oscillation phase is $\delta m^2_{\rm mat} t_{\rm sc}/(4E) \ll 1$.  In this case, the flavor change probability for $T \geq T_{\rm eq}$ is modified to
\begin{eqnarray}
  P_{\rm as} &\approx& \sin^2 2 \tilde{\theta} \sin^2 \left(\frac{\delta m^2_{\rm mat} t_{\rm sc}}{4E} \right)^2
  \simeq \sin^2 2\theta
  {\left( \frac{\delta m^2}{G_{\rm F}^2 T^6} \right)^2} \nonumber \\
  &=& \sin^2 2\theta
    \left( \frac{T}{T_{\rm eq}} \right)^{-12}
    ~~~({\rm for}~ T \geq T_{\rm eq}).
    \label{eq_ap19}
  \end{eqnarray}
We find that the flavor change probability is smaller at high temperatures since there is no enough time for oscillation.  We note that this probability scales similarly to that of the 3D space case with $\epsilon_{\rm s} =0$ (see Eqs. (\ref{eq_ap1}) and (\ref{eq_ap18})).  We thus confirm that the resonant extra-dimensional effect possibly increases the effective mixing angle while it simultaneously increases the oscillation time scale, i.e., $t_{\rm osc} =4E/\delta m^2_{\rm mat}$.  As a result, the flavor change probability is not changed drastically from that of the standard three dimensional case.

\section{Solution of the Boltzmann equation}\label{appendix2}

In this paper, we utilized the rate equation instead of the Boltzmann equation in the estimation of the relic energy density of sterile neutrino.
We check how well the result of the rate equation approximates the exact result.

\begin{widetext}
\subsection{$\nu$-oscillation in the universe}\label{appendix2_1}
Before the decoupling of active neutrinos, the neutrino oscillation phase is given by
\begin{eqnarray}
  \alpha_{\rm osc} &=& \frac{\delta m_{\rm mat}^2 t_{\rm sc}}{4E} \\
  &=& 3.8 \times 10^{8} \left( \frac{\delta m_{\rm mat}^2}{{\rm eV}^2} \right)
  \left( \frac{t_{\rm sc}}{\rm s} \right)
  \left( \frac{E}{\rm MeV} \right)^{-1}.
\end{eqnarray}

At the beginning of BBN of $t =$1 s and $T =$ 1 MeV, the neutrino oscillation of $\nu_{\rm a}$ and $\nu_{\rm s}$ is very frequent on the cosmic expansion time scale for $\delta m^2 ~^>_\sim 1$ eV$^2$.  In the early universe of $T >$ 1 MeV, this phase is usually larger than unity [Eq. (\ref{eq_ap17})]. For simplicity, we assume a case in which the flavor change probability is given by Eq. (\ref{eq_ap18}).

The production rate of $\nu_{\rm s}$ is given by
\begin{equation}
  \Gamma_{\nu_{\rm s}}(E) =\frac{1}{2} \sin^2 ( 2 \tilde{\theta}(E)) \Gamma_{\rm W}(E),
\end{equation}
where the factor $\sin^2 ( 2 \tilde{\theta}) /2$ is the probability of flavor change from $\nu_{\rm a}$ to $\nu_{\rm s}$ after the production of $\nu_{\rm a}$, and $\Gamma_{\rm W}$ is the rate of weak reaction which produces $\nu_{\rm a}$.

When the value of $\tilde{\theta}$ is large, i.e., $\tilde{\theta} ~^<_\sim 1$, the flavor change becomes maximally effective.
\begin{enumerate}
\item If this effective epoch is before the freeze-out of $\nu_{\rm a}$, $\Gamma_{\nu_{\rm s}} >H^{-1}$ is realized.  Then, the $\nu_{\rm s}$ abundance approaches to the equilibrium value.
  \item If this effective epoch is after the freeze-out of $\nu_{\rm a}$, the oscillation leads to an equalization of energy densities for $\nu_{\rm s}$ and $\nu_{\rm a}$.  Because of the energy conservation, however, the total neutrino energy density is unchanged.  Therefore, the additional neutrino energy, i.e., $\Delta \rho_\nu$, is not affected.

\end{enumerate}

We assume that the mass squared difference is larger than $\delta m^2 \sim$ eV$^2$ as considered in Pas et al. (2005).  There are constraints on the mixing angle, f.e., $\sin^2 2\tilde{\theta}_{24} \lesssim 10^{-1}$ for $\delta m_{41}^2 \sim $1 eV$^2$ (IceCube) \cite{TheIceCube:2016oqi} and $|U_{\mu 4}|^2 <0.041$ and $|U_{\tau 4}|^2 <0.18$ for $\delta m^2 >0.1$ eV$^2$ (90 \% C.L.) (Super-Kamiokande) \cite{Abe:2014gda}.  We then assume that the bare mixing angle $\theta$ is significantly smaller than unity in this case.

\subsection{Upper limit on the mass}\label{appendix2_2}

We focus on the relatively heavy sterile neutrino case, and consider an upper limit on the mass. If the sterile neutrino can decay before the active neutrino decoupling, there is no sterile neutrino in BBN epoch.  The $\nu_{\rm s}$ mass can then be constrained from the requirement of $\tau > 1$ s in order to have any effect on BBN.  The decay rate is given [Eq. (7.12) in Ref. \cite{Ishida:2014wqa}] by
\begin{equation}
  \Gamma_{\rm dec} \sim 1.87 \times 10^{-5}~{\rm s}^{-1}
  \left( \frac{\tilde{\theta}}{10^{-3}}\right)^2
  \left( \frac{m_{\nu_{\rm s}}}{14~{\rm MeV}} \right)^5.
\end{equation}
The condition of $\Gamma_{\rm dec} =\tau^{-1} < 1$ s$^{-1}$ is then satisfied when
\begin{equation}
  m_{\nu_{\rm s}} < 20~{\rm MeV} \left( \frac{\tilde{\theta}}{0.1}\right)^{-2/5}.
\end{equation}

\subsection{Full width at 1/e maximum of the resonance}\label{appendix2_3}
A maximum in the effective mixing angle as a function of $E$ is derived as follows:
We define the function
\begin{eqnarray}
  f(E; T) &=& \sin^2 2 \tilde{\theta} \nonumber \\
  &=& \frac{\sin^2 2 \theta}{\sin^2 2 \theta +\cos^2 2 \theta \left[ 1 +F(E_{\rm res}, T) E^2 \right]^2},
\end{eqnarray}
where we defined
\begin{eqnarray}
  F(E_{\rm res}, T) &=& D(T) -1/E_{\rm res}^2,
  \label{eq_full8} \\
D(T) &=& \frac{C_\alpha G_{\rm F}^2 T^4}{\cos 2 \theta \alpha \delta m^2}.
\end{eqnarray}
In these equations, $\theta$ is the bare mixing angle between the sterile and active neutrinos, $E_{\rm res}$ is a parameter related to the extra-dimension [Eq. (\ref{eq16})],
$C_\alpha$ is the flavor ($\alpha$) dependent constant,
$G_{\rm F}$ is the Fermi constant,
$\alpha$ is the fine structure constant, and
$\delta m^2$ is the mass squared difference of the sterile and active neutrinos.

The derivative of this function with respect to $E$ is given by
\begin{equation}
  \frac{df(E; T)}{dE} = \frac{-4\sin^2 2 \theta \cos^2 2 \theta
  \left[ 1 +F(E_{\rm res}, T) E^2 \right]
       F(E_{\rm res}, T) E}
       {\left\{\sin^2 2 \theta +\cos^2 2 \theta \left[ 1 +F(E_{\rm res}, T) E^2 \right]^2 \right\}^2}.
\end{equation}
Maxima exist for $df/dE=0$, i.e.,
\begin{eqnarray}
  \left[ 1 +F(E_{\rm res}, T) E^2 \right] F(E_{\rm res}, T) =0.
\end{eqnarray}

\begin{enumerate}
\item $F(E_{\rm res}, T) =0$ case

  When $F(E_{\rm res}, T) =0$ is satisfied, $D(T) =1/E_{\rm res}^2$ is hold.  In this case, the effective mixing angle is the same as the mixing angle $\theta$ independent of $E$.  Therefore, $f(E; T)$ is constant, and the condition $df/dE =0$ is realized for any $E$.  Thus, at the temperature satisfying $D(T) =1/E_{\rm res}^2$, there is no maximum.  This temperature occurs only once.

\item $E^2 =-1 /F(E_{\rm res}, T)$ case

  For this case, the peak energy is given by
  \begin{equation}
    E_{\rm peak}(E_{\rm res}, T) = \frac{E_{\rm res}}{\sqrt{\mathstrut 1 -D(T) E_{\rm res}^2}}.
    \label{eq_full12}
  \end{equation}
  The maximum is given by
  \begin{equation}
    f(E_{\rm peak}; T) = \frac{\sin^2 2 \theta}{\sin^2 2 \theta +\cos^2 2 \theta \left[ 1 +F(E_{\rm res}, T) E_{\rm peak}^2 \right]^2}
    =1.
\end{equation}
  This maxima exist only for $F(E_{\rm res}, T)<0$, i.e., $D(T) < 1/E_{\rm res}^2$.  Therefore, the resonance appears after the temperature of the universe decreases to some critical temperature.

  After the condition $D(T) E_{\rm res}^2=1$ is satisfied, the peak energy quickly moves from infinity to $E_{\rm res}$ as the temperature decreases. The asymptotic value of $E_{\rm peak}$ at low $T$ values is $E_{\rm res}$. We note that sterile neutrinos with energies below a critical value do not experience any resonance since this peak energy never overlaps the sterile neutrino energy. On the other hand, sterile neutrino with energies above the critical value have two resonance epochs in general. The sterile neutrino energy that is red-shifting once becomes larger than $E_{\rm peak}$ during the peak energy is decreasing. After that, the red-shifting energy becomes smaller than $E_{\rm peak} \approx E_{\rm res}$. These behaviors of the first and second resonances are shown in Appendix \ref{appendix2_6} below.

The full width at 1/e maximum of $\sin^2 2 \tilde{\theta}(E)$ (for $F<0$) is derived as follows:
\begin{eqnarray}
  \sin^2 2 \tilde{\theta} = \frac{\sin^2 2 \theta}{\sin^2 2 \theta +\cos^2 2 \theta \left[ 1 + F(E_{\rm res}, T) E^2 \right]^2} &\geq& \frac{1}{e} \\
  \Longrightarrow~~~ \sqrt{\mathstrut 1 -\tan 2 \theta \sqrt{\mathstrut e-1}} &\leq& \frac{E}{E_{\rm peak}} \leq \sqrt{\mathstrut 1 +\tan 2 \theta \sqrt{\mathstrut e-1}} \\
  \Longrightarrow~~~ 1 - \theta \sqrt{\mathstrut e-1} &\leq& \frac{E}{E_{\rm peak}} \leq
  1 + \theta \sqrt{\mathstrut e-1}~~({\rm for}~\theta \ll 1).
\end{eqnarray}

We note that the energy at which the maximum of the function $f$ appears, i.e., $E_{\rm peak}$, depends on $T$. Especially, at the second resonance, the matter term $D(T) E_{\rm res}^2$ in Eq. (\ref{eq_full12}) is subdominant, and the energy $E_{\rm peak}$ does not significantly dependent on $T$.  When we approximate $E_{\rm peak}$ with $E_{\rm res}$, the full width at 1/e maximum is given by
\begin{eqnarray}
  \frac{\Delta E}{E_{\rm res}} &\approx &
  2 \theta \sqrt{\mathstrut e-1}~~({\rm for}~\theta \ll 1)
  \label{eq_b16} \\
  \Longrightarrow~~~ \Delta \ln a &=&
  2 \theta \sqrt{\mathstrut e-1}~~({\rm for}~\theta \ll 1),
  \label{eq_b17}
\end{eqnarray}
where $\Delta \ln a$ is the scale factor interval in logarithmic scale corresponding to the duration of the second resonance of the mixing angle.

\end{enumerate}

\subsection{Boltzmann equation}\label{appendix2_4}
As the neutrino energy redshifts, it pass through the resonant region in the effective mixing angle $\tilde{\theta}(E)$.  Although the width of resonance can be narrow, when $T\sim E_{\rm res}$ is satisfied, all energy region of $E\sim T$ experiences the resonance peak.  Therefore, the approximation of Boltzmann equation by the rate equation would not introduce a very large error in the final sterile neutrino abundance although there is certainly some error.

The Boltzmann equation of the sterile neutrino in the Friedmann-Lema\^{i}tre-Robertson-Walker Universe is given \cite{Kolb1994,Dolgov:2000pj} by

\begin{equation}
  \left( \partial_t -H p \partial_p \right) f_k(p,t)
  = H a \partial_a f_k(y,a) =I_{\rm coll},
  \label{eq_bol_1}
\end{equation}
where
\begin{eqnarray}
  I_{\rm coll} =\frac{1}{2E_k} \sum_{\rm process} \int
  \prod_{i\neq k} \left[ \frac{d^3 p_i}{2E_i\left(2 \pi \right)^3} \right]
  \prod_{f\neq k} \left[ \frac{d^3 p_f}{2E_f\left(2 \pi \right)^3} \right]
  \left( 2\pi \right)^4 \delta^{(4)} \left( \sum_i p_i -\sum_f p_f \right)
  \frac{1}{2} S \left| A_{if} \right|^2
  F(f_i,f_f)
  \label{eq_bol_2}
\end{eqnarray}
is the collision integral with
\begin{equation}
  F(f_i,f_f) = -\prod_i f_i \prod_f \left(1-f_f \right)
  +\prod_f f_f \prod_i \left(1-f_i \right)
  \label{eq_bol_3}
\end{equation}
the factor for the phase space.
In these equations,
$t$ is the cosmic time,
$p$ is the momentum,
$H=\dot{a}/a$ is the cosmic expansion rate
with $a(t)$ the scale factor of the universe,
$f_l$ is the phase space distribution function of a fermion $l$,
$E_l$ is the total energy of $l$,
indexes $i$ and $j$ are used for particles in the initial and final states, respectively,
the factor of 1/2 is for taking a spin average for particles in the initial state,
$S=1/m!$ with $m$ the number of identical particles in the final state, and
$A_{ij}$ is the matrix element.
In Eq. (\ref{eq_bol_2}), the sum is taken over process.
At the first equality in Eq. (\ref{eq_bol_1}), the variable $y=p a(t)$ is defined and the distribution function $f(a,y)$ is considered.

Matrix elements of a sterile neutrino are listed in Tables 1 and 2 in Ref. \cite{Dolgov:2000pj}.  We consider the relatively light sterile neutrino, i.e., $m_{\nu_{\rm s}} < 2 m_e$, where $m_e =0.510999$ MeV is the electron mass.  Then, the sterile neutrino decay into an $e^+e^-$ pair and an active neutrino does not occur energetically.  In addition, since we consider cosmic temperatures which are well above the electron mass, terms proportional to $m_e^2$ in matrix elements can be neglected.  Furthermore, it is assumed that all fermions excepting the sterile neutrino have the exact Fermi-Dirac distribution and that masses of those fermions are neglected.  In the decay and scattering processes, we adopt indexes as $1\rightarrow 2+3+4$ and $1+2 \rightarrow 3+4$ and identify the index 1 to be the sterile neutrino.

The collision terms for the decay and scattering are then given respectively by
\begin{eqnarray}
  I_{\rm coll,d} =\frac{4}{(2\pi)^5} G_{\rm F}^2 \tilde{\theta}^2
  \frac{1}{E_1} \int \frac{d^3 p_2}{E_2} \frac{d^3 p_3}{E_3} \frac{d^3 p_4}{E_4}
    \delta^{(4)} \left[ p_1 -\left(p_2 +p_3 +p_4 \right) \right]
  (p_1 \cdot p_4) (p_2 \cdot p_3) F(f_i,f_f),
  \label{eq_bol_4}
\end{eqnarray}
and
\begin{eqnarray}
  I_{\rm coll,s} &=& \frac{4\left(1+\tilde{g}_{\rm L}^2 +g_{\rm R}^2 \right)}
  {(2\pi)^5} G_{\rm F}^2 \tilde{\theta}^2
  \frac{1}{E_1} \int \frac{d^3 p_2}{E_2} \frac{d^3 p_3}{E_3} \frac{d^3 p_4}{E_4}
    \delta^{(4)} \left[ p_1 +p_2 -\left(p_3 +p_4 \right) \right] \nonumber \\
    && \hspace{13em} \times \left[2(p_1 \cdot p_4) (p_2 \cdot p_3) +(p_1 \cdot p_2) (p_3 \cdot p_4) \right]
    F(f_i,f_f),
  \label{eq_bol_5}
\end{eqnarray}
where
we defined
\begin{eqnarray}
  \tilde{g}_{\rm L} &=& -\frac{1}{2} +\sin^2 \theta_{\rm W},
  \label{eq_bol_6} \\
  g_{\rm R} &=& \sin^2 \theta_{\rm W},
  \label{eq_bol_7}
\end{eqnarray}
with the weak angle $\sin^2 \theta_{\rm W}=0.23$ \cite{Agashe:2014kda}.

The two terms in Eq. (\ref{eq_bol_5}) are separately defined as
\begin{eqnarray}
  I_{\rm coll,s}^{(1)} &=& \frac{4\left(1+\tilde{g}_{\rm L}^2 +g_{\rm R}^2 \right)}
  {(2\pi)^5} G_{\rm F}^2 \tilde{\theta}^2
  \frac{1}{E_1} \int \frac{d^3 p_2}{E_2} \frac{d^3 p_3}{E_3} \frac{d^3 p_4}{E_4}
    \delta^{(4)} \left[ p_1 +p_2 -\left(p_3 +p_4 \right) \right] \nonumber \\
    && \hspace{13em} \times 2(p_1 \cdot p_4) (p_2 \cdot p_3)
    F(f_i,f_f)
    \label{eq_bol_8} \\
  I_{\rm coll,s}^{(2)} &=& \frac{4\left(1+\tilde{g}_{\rm L}^2 +g_{\rm R}^2 \right)}
  {(2\pi)^5} G_{\rm F}^2 \tilde{\theta}^2
  \frac{1}{E_1} \int \frac{d^3 p_2}{E_2} \frac{d^3 p_3}{E_3} \frac{d^3 p_4}{E_4}
    \delta^{(4)} \left[ p_1 +p_2 -\left(p_3 +p_4 \right) \right] \nonumber \\
    && \hspace{13em} \times (p_1 \cdot p_2) (p_3 \cdot p_4)
    F(f_i,f_f).
    \label{eq_bol_9}
\end{eqnarray}

Performing the integrals, we obtain the exact formulae for $\mbox{\boldmath $p_1$}\neq 0$ as follows:  For the decay term,
\begin{eqnarray}
  I_{\rm coll,d} &=&\frac{1}{2\pi^3} G_{\rm F}^2 \tilde{\theta}^2
  \left\{
  -\frac{f_1(E_1)}{p_1}
  \left[ \int_0^{\frac{E_1 -p_1}{2}} dE_4 \int_{|p_1 -p_4|}^{p_1+p_4} dR
    +\int_{\frac{E_1 -p_1}{2}}^{\frac{E_1 +p_1}{2}} dE_4 \int_{|p_1 -p_4|}^{E_1-E_4} dR
    \right] \right. \nonumber \\
  && \left. \times \left[ 1-f_4(E_4) \right]
  \left[ \left(E_1 -E_4 \right)^2 -R^2 \right]
  \left[ p_4 -\frac{p_1^2 +p_4^2 -R^2}{2E_1} \right]
  G_1(E_1,E_4,R) \right. \nonumber \\
  && \left. +\frac{1-f_1(E_1)}{p_1}
\left[ \int_0^{\frac{E_1 -p_1}{2}} dE_4 \int_{|p_1 -p_4|}^{p_1+p_4} dR
    +\int_{\frac{E_1 -p_1}{2}}^{\frac{E_1 +p_1}{2}} dE_4 \int_{|p_1 -p_4|}^{E_1-E_4} dR
    \right] \right. \nonumber \\
  && \left. f_4(E_4)
  \left[ \left(E_1 -E_4 \right)^2 -R^2 \right]
  \left[ p_4 -\frac{p_1^2 +p_4^2 -R^2}{2E_1} \right]
  G_2(E_1,E_4,R) \right\}
  \label{eq_bol_10} \\
  G_1(E_1,E_4,R) &\equiv& \int_{E_{2{\rm min,d}}}^{E_{2{\rm max,d}}} \left[ 1-f_2(E_2)\right]
  \left[ 1-f_3(E_1-E_4-E_2)\right] dE_2
  \label{eq_bol_11} \\
    &=&
    \begin{cases}
      \frac{T}{1-e^{-a_2}}
      \left[\ln \frac{\exp(x_{2{\rm max,d}}) +1}{\exp(x_{2{\rm max,d}}) +\exp(a_2)}
        -\ln \frac{\exp(x_{2{\rm min,d}}) +1}{\exp(x_{2{\rm min,d}}) +\exp(a_2)} \right] & (a_2\neq 0) \\
      T \left( \frac{1}{\exp(x_{2{\rm min,d}}) +1} -\frac{1}{\exp(x_{2{\rm max,d}}) +1} \right) & (a_2 =0) \\
    \end{cases}
    \label{eq_bol_12} \\
  G_2(E_1,E_4,R) &\equiv& \int_{E_{2{\rm min,d}}}^{E_{2{\rm max,d}}} f_2(E_2)
  f_3(E_1-E_4-E_2) dE_2
  \label{eq_bol_13} \\
    &=&
    \begin{cases}
      \frac{T}{e^{a_2} -1}
      \left[\ln \frac{\exp(x_{2{\rm max,d}}) +1}{\exp(x_{2{\rm max,d}}) +\exp(a_2)}
        -\ln \frac{\exp(x_{2{\rm min,d}}) +1}{\exp(x_{2{\rm min,d}}) +\exp(a_2)} \right] & (a_2\neq 0) \\
      T \left( \frac{1}{\exp(x_{2{\rm min,d}}) +1} -\frac{1}{\exp(x_{2{\rm max,d}}) +1} \right) & (a_2 =0) \\
    \end{cases}
    \label{eq_bol_14} \\
    x_{2{\rm min,d}} &=&\frac{E_{2{\rm min,d}}}{T} =\frac{E_1 -E_4 -R}{2T}
    \label{eq_bol_15} \\
    x_{2{\rm max,d}} &=&\frac{E_{2{\rm max,d}}}{T} =\frac{E_1 -E_4 +R}{2T}
    \label{eq_bol_16} \\
    a_2&=& \frac{E_1 -E_4}{T}.
    \label{eq_bol_17}
\end{eqnarray}
For scattering terms, we obtain
\begin{eqnarray}
  I_{\rm coll,s}^{(1)} &=& \frac{\left(1+\tilde{g}_{\rm L}^2 +g_{\rm R}^2 \right)}
  {2\pi^3} G_{\rm F}^2 \tilde{\theta}^2
  \left\{
  -\frac{f_1(E_1)}{p_1}
  \left[ \int_{\frac{E_1 -p_1}{2}}^{\frac{E_1 +p_1}{2}} dE_4 \int_{E_1 -E_4}^{p_1+E_4} dR
    +\int_{\frac{E_1 +p_1}{2}}^\infty dE_4 \int_{E_4 -p_1}^{E_4+p_1} dR
    \right] \right. \nonumber \\
  && \left. \times \left[ 1-f_4(E_4) \right]
  \left[ R^2 -\left(E_1 -E_4 \right)^2 \right]
  \left[ p_4 -\frac{p_1^2 +p_4^2 -R^2}{2E_1} \right]
  G_3(E_1,E_4,R) \right. \nonumber \\
  && \left. +\frac{1-f_1(E_1)}{p_1}
  \left[ \int_{\frac{E_1 -p_1}{2}}^{\frac{E_1 +p_1}{2}} dE_4 \int_{E_1 -E_4}^{p_1+E_4} dR
    +\int_{\frac{E_1 +p_1}{2}}^\infty dE_4 \int_{E_4 -p_1}^{E_4+p_1} dR
    \right] \right. \nonumber \\
  && \left. f_4(E_4)
  \left[ R^2 -\left(E_1 -E_4 \right)^2 \right]
  \left[ p_4 -\frac{p_1^2 +p_4^2 -R^2}{2E_1} \right]
  G_4(E_1,E_4,R) \right\}
  \label{eq_bol_18} \\
  G_3(E_1,E_4,R) &\equiv& \int_{E_{2{\rm min,s}}}^\infty f_2(E_2)
  \left[ 1-f_3(E_1-E_4+E_2)\right] dE_2
  \label{eq_bol_19} \\
    &=&
    \begin{cases}
      \frac{T}{1-e^{-a_2}}
      \left[a_2-\ln \frac{\exp(x_{2{\rm min,s}} +a_2) +1}{\exp(x_{2{\rm min,s}}) +1} \right]
      & (a_2\neq 0) \\
      T \frac{1}{\exp(x_{2{\rm min,s}}) +1} & (a_2 = 0) \\
    \end{cases}
    \label{eq_bol_20} \\
    G_4(E_1,E_4,R) &\equiv& \int_{E_{2{\rm min,s}}}^\infty \left[ 1- f_2(E_2) \right]
    f_3(E_1-E_4+E_2) dE_2
    \label{eq_bol_21} \\
    &=&
    \begin{cases}
      \frac{T}{e^{a_2} -1}
      \left[a_2 -\ln \frac{\exp(x_{2{\rm min,s}} +a_2) +1}{\exp(x_{2{\rm min,s}}) +1} \right]
      & (a_2\neq 0) \\
      T \frac{1}{\exp(x_{2{\rm min,s}}) +1} & (a_2 = 0) \\
    \end{cases}
    \label{eq_bol_22} \\
    x_{2{\rm min,s}} &=&\frac{E_{2{\rm min,s}}}{T} =\frac{R -E_1 +E_4}{2T}
    \label{eq_bol_23} \\
    a_2&=& \frac{E_1 -E_4}{T},
    \label{eq_bol_24}
\end{eqnarray}
and
\begin{eqnarray}
  I_{\rm coll,s}^{(2)} &=& \frac{\left(1+\tilde{g}_{\rm L}^2 +g_{\rm R}^2 \right)}
  {4\pi^3} G_{\rm F}^2 \tilde{\theta}^2
  \left\{
  -\frac{f_1(E_1)}{p_1}
  \int_0^\infty f_2(E_2) dE_2 \int_{|p_1 -p_2|}^{p_1+p_2} dR
  \right. \nonumber \\
  && \left. \times
  \left[ \left(E_1 +E_2 \right)^2 -R^2 \right]
  \left[ p_2 -\frac{R^2 -p_1^2 -p_2^2}{2E_1} \right]
  G_5(E_1,E_2,R) \right. \nonumber \\
  && \left. +\frac{1-f_1(E_1)}{p_1}
  \int_0^\infty \left[ 1- f_2(E_2) \right] dE_2 \int_{|p_1 -p_2|}^{p_1+p_2} dR
  \right. \nonumber \\
  && \left. \times
  \left[ \left(E_1 +E_2 \right)^2 -R^2 \right]
  \left[ p_2 -\frac{R^2 -p_1^2 -p_2^2}{2E_1} \right]
  G_6(E_1,E_2,R) \right\}
  \label{eq_bol_25} \\
  G_5(E_1,E_2,R) &\equiv& \int_{E_{4{\rm min,s}}}^{E_{4{\rm max,s}}}
  \left[ 1-f_4(E_4)\right]
  \left[ 1-f_3(E_1+E_2 -E_4)\right] dE_4
  \label{eq_bol_26} \\
  &=&
  \begin{cases}
    \frac{T}{1-e^{-a_4}}
    \left[ \ln \frac{\exp(x_{4{\rm max,s}}) +1}{\exp(x_{4{\rm max,s}}) +\exp(a_4)}
      -\ln \frac{\exp(x_{4{\rm min,s}}) +1}{\exp(x_{4{\rm min,s}}) +\exp(a_4)} \right]
    & (a_4\neq 0) \\
    T \left( \frac{1}{\exp(x_{4{\rm min,s}}) +1} -\frac{1}{\exp(x_{4{\rm max,s}}) +1} \right) & (a_4 = 0) \\
  \end{cases}
  \label{eq_bol_27} \\
  G_6(E_1,E_2,R) &\equiv& \int_{E_{4{\rm min,s}}}^{E_{4{\rm max,s}}}
  f_4(E_4)
  f_3(E_1+E_2 -E_4) dE_4
  \label{eq_bol_28} \\
    &=&
  \begin{cases}
    \frac{T}{e^{a_4} -1}
    \left[ \ln \frac{\exp(x_{4{\rm max,s}}) +1}{\exp(x_{4{\rm max,s}}) +\exp(a_4)}
      -\ln \frac{\exp(x_{4{\rm min,s}}) +1}{\exp(x_{4{\rm min,s}}) +\exp(a_4)} \right]
    & (a_4\neq 0) \\
    T \left( \frac{1}{\exp(x_{4{\rm min,s}}) +1} -\frac{1}{\exp(x_{4{\rm max,s}}) +1} \right) & (a_4 = 0) \\
  \end{cases}
  \label{eq_bol_29} \\
  x_{4{\rm min,s}} &=&\frac{E_{4{\rm min,s}}}{T} =\frac{E_1 +E_2 -R}{2T}
  \label{eq_bol_30} \\
  x_{4{\rm max,s}} &=&\frac{E_{4{\rm max,s}}}{T} =\frac{E_1 +E_2 +R}{2T}
  \label{eq_bol_31} \\
  a_4&=& \frac{E_1 +E_2}{T}.
  \label{eq_bol_32}
\end{eqnarray}

We note that in this formulation, we adopted variables $\mbox{\boldmath $R$}= \mbox{\boldmath $p_1$}-\mbox{\boldmath $p_4$}$ (for terms $I_{\rm coll,d}$ and $I_{\rm coll,s}^{(1)}$) and $\mbox{\boldmath $R$}= \mbox{\boldmath $p_1$}+\mbox{\boldmath $p_2$}$ (for a term $I_{\rm coll,s}^{(2)}$).

Especially, when the mass of the sterile neutrino is much larger than the temperature, the Pauli blocking effect is negligible in the phase factor [Eq. (\ref{eq_bol_3})].  Then, the first term in Eq. (\ref{eq_bol_10}) becomes
\begin{eqnarray}
{\rm 1st~term~of~} I_{\rm coll,d} &=& -\frac{1}{192\pi^3}
G_{\rm F}^2 \tilde{\theta}^2 m_{\nu_{\rm s}}^5
f_1(E_1).
\label{eq_bol_33}
\end{eqnarray}
This gives the life time of sterile neutrino at low temperatures, i.e.,
\begin{eqnarray}
  \tau_{\nu_s}(T=0) =  \left[ \frac{1}{192\pi^3}
  G_{\rm F}^2 \tilde{\theta}^2 m_{\nu_{\rm s}}^5 \right]^{-1}.
  \label{eq_bol_34}
\end{eqnarray}

By using replacement for terms of distribution function as $(1 -f_l) \leftrightarrow 1$ and $f_l =[\exp(E_l/T) +1]^{-1} \leftrightarrow \exp(-E_l/T)$, inaccurate and analytic expressions for the collision terms are derived and used frequently.  For example, Eq. (23) in Ref. \cite{Dolgov:2000pj} for $p_{\nu_{\rm s}}=0$ ($E_{\nu_{\rm s}} =m_{\nu_{\rm s}}$) is reproduced using the replacement in Eqs. (\ref{eq_bol_4}) and (\ref{eq_bol_5}).  However, an error of a factor of up to two is introduced by each replacement of $(1 -f_l) \leftrightarrow 1$ or $f_l =[\exp(E_l/T) +1]^{-1} \leftrightarrow \exp(-E_l/T)$, in general.  Therefore, we should use the exact collision terms as given above.

\subsection{Abundance increase at the resonance}\label{appendix2_5}
We assume that the sterile neutrino is ultrarelativistic before the decoupling.  The equilibrium distribution function of fermion, i.e., the Fermi-Dirac function, is given by
\begin{equation}
  f_{\rm EQ}(E, t) =\frac{1}{\exp(E/T(t)) +1},
  \label{eq_dis_1}
\end{equation}
\begin{equation}
  f_{\rm EQ}(y,a) =\frac{1}{\exp\left\{y/[aT(a)]\right\} +1},
  \label{eq_dis_2}
\end{equation}
where
we define $y\equiv E_0$ and $a_0=1$ as the sterile neutrino energy and the scale factor at the initial temperature $T_0=100$ GeV.
The product $aT$ has the scaling derived hereinbelow.

Suppose that the abundance of the sterile neutrino is very small initially and it increases significantly during the resonance epoch.  If the final abundance does not reach the equilibrium abundance, the abundance change roughly scales as
\begin{eqnarray}
  \Delta f(y,a)_{\rm res} &\sim& \Gamma_{\nu_{\rm s}}(E_{\rm peak}, T_{\rm peak}(y)) \Delta t_{\rm peak} f_{\rm EQ}(y,a_{\rm peak}) \nonumber \\
  &\propto& \Gamma_{\nu_{\rm s}}(E_{\rm peak}, T_{\rm peak}(y)) \theta \frac{1}{H(T_{\rm peak}(y))} f_{\rm EQ}(y, a_{\rm peak}) \nonumber \\
  &\propto& \theta\, T_{\rm peak}(y)^3 f_{\rm EQ}(y, a_{\rm peak}),
  \label{eq_dis_3}
\end{eqnarray}
where
$T_{\rm peak}(y)$ is the temperature at which the resonant mixing occurs for a given $y$,
$a_{\rm peak}(y)$ is the scale factor corresponding to the temperature, and
$\Gamma_{\nu_{\rm s}}(E_{\rm peak}, T_{\rm peak}(y))$ is the sterile neutrino production rate at the energy $E_{\rm peak}$ and the temperature $T_{\rm peak}(y)$.

On the other hand, if the reaction rate is very large, the final abundance becomes the equilibrium abundance.  Since the difference in the final abundance between the exact and approximated treatment using the Boltzmann and the rate equations, respectively, is small in the latter case, we focus on the former case in what follows.

The resonant mixing for a fixed $y$ value occurs when the energy redshifts to the peak energy, i.e.,
\begin{eqnarray}
  E_{\rm peak} =\frac{y}{a_{\rm peak}(y)}.
  \label{eq_dis_4}
\end{eqnarray}

The entropy per comoving volume is given by
\begin{equation}
  S = s a^3 = \frac{2 \pi^2}{45} g_{\ast {\rm S}} T^3 a^3,
  \label{eq_dis_5}
\end{equation}
where
$s$ is the entropy density of the universe, and
$g_{\ast {\rm S}}$ is the statistical degrees of freedom for entropy.
The entropy conservation during the resonance leads to the equation
\begin{equation}
  T_{\rm peak}(y) =\frac{T_0}{a_{\rm peak}(y)}  \left( \frac{g_{\ast {\rm S}0}}{g_{\ast {\rm S},{\rm peak}}(y)} \right)^{1/3}
  =T_0 \frac{E_{\rm peak}}{y}  \left( \frac{g_{\ast {\rm S}0}}{g_{\ast {\rm S},{\rm peak}}(y)} \right)^{1/3},
  \label{eq_dis_6}
\end{equation}
where
$g_{\ast {\rm S}0}$ and $g_{\ast {\rm S},{\rm peak}}(y)$ are the values of $g_{\ast {\rm S}}$ at $T_0$ and $T_{\rm peak}(y)$, respectively.

We then obtain
\begin{equation}
  f_{\rm EQ}(E_{\rm peak}, t(a_{\rm peak}(y))) =\frac{1}{\exp(E_{\rm peak}/T_{\rm peak}(y)) +1}
  =\frac{1}{\exp \left[\frac{y}{T_0} \left( \frac{g_{\ast {\rm S},{\rm peak}}(y)}{g_{\ast {\rm S}0}} \right)^{1/3} \right] +1}.
  \label{eq_dis_7}
\end{equation}

The change in distribution function is approximately given by
\begin{eqnarray}
  \Delta f(y,a)_{\rm res} &\propto& \frac{\theta T_{\rm peak}(y)^3}{\exp({E_{\rm peak}/T_{\rm peak}}) +1}
  \label{eq_dis_8}.
\end{eqnarray}
At the first resonance, the peak temperature is rather constant [see Eq. (\ref{eq_b65}) below] since the value of $E_{\rm peak}$ quickly evolves. The change is then given by
\begin{eqnarray}
  \Delta f(y,a)_{\rm res}^1 &\propto& \frac{1}
         {\exp \left[\frac{y}{T_0} \left( \frac{g_{\ast {\rm S},{\rm peak}}(y)}{g_{\ast {\rm S}0}} \right)^{1/3} \right] +1}.
         \label{eq_dis_9}
\end{eqnarray}
At the second resonance, on the other hand, the peak energy is close to $E_{\rm res}$ and the value of $T_{\rm peak}(y)$ significantly depends on $y$. The change is then given by
\begin{eqnarray}
  \Delta f(y,a)_{\rm res}^2 &\propto& \frac{1}
{y^3 g_{\ast {\rm S},{\rm peak}}(y) } \frac{1}{\exp \left[\frac{y}{T_0} \left( \frac{g_{\ast {\rm S},{\rm peak}}(y)}{g_{\ast {\rm S}0}} \right)^{1/3} \right] +1}.
  \label{eq_dis_10}
\end{eqnarray}

\subsection{Test calculation}\label{appendix2_6}
We check a difference in the distribution function of the sterile neutrino derived from the exact calculation and the simplified estimation.  In order to check the expectable maximum difference, we choose a case where the initial abundance of the sterile neutrino is negligible.  For example, we take $m_{\nu_{\rm s}} =1$ eV, $\theta=10^{-8}$, and $E_{\rm res}=10$ MeV.  Then, even at the cosmic temperature of the electro-weak phase transition of $T\sim 200$ GeV, the sterile neutrino is not in the equilibrium.

This is shown by the fact that the sterile neutrino production rate is smaller than the cosmic expansion rate using the following equation:
The production rate of the sterile neutrino and the cosmic expansion rate are respectively given [Eqs. (\ref{eq_ap10}) and (\ref{eq_ap18})] by
\begin{eqnarray}
  \Gamma_{\nu_{\rm s}} &\sim& G_{\rm F}^2 \tilde{\theta}^2 T^5, \\
  H &\sim & \frac{g_\ast^{1/2} T^2}{M_{\rm Pl}},
\end{eqnarray}
where
$g_\ast$ is the statistical degrees of freedom for energy, and
$M_{\rm Pl}$ is the Planck mass.
Then, we have a relation [Eq. (7.11) in Ref. \cite{Ishida:2014wqa}]:
\begin{eqnarray}
  \frac{\Gamma_{\nu_{\rm s}}}{H} &\sim& \left( \frac{\tilde{\theta}}{10^{-3}}\right)^2
  \left( \frac{g_\ast}{63.75} \right)^{-1/2}
  \left( \frac{T}{0.2~{\rm GeV}} \right)^3.
\end{eqnarray}

For the adopted parameter set, the sterile neutrino abundance is very small before the resonant mixing occurs.  Therefore, we can assume that the abundance is zero at the initial time of the calculation.  We can then estimate the maximum difference in the distribution function calculated by the Boltzmann equation and the rate equation from this result.  We note that the flavor change probability is the average value for the case of complete oscillation [Eq. (\ref{eq_ap18})] in the whole temperature region until the sterile neutrino decoupling for this parameter set (see Appendix \ref{appendix1}).

Figure \ref{fig_b1} shows the ratio of the calculated distribution function and the equilibrium function, i.e., $f/f_{\rm EQ} (y)$, (solid lines) as a function of temperature for $y/T_0=0.25$, $1$, $2$, $3.15$, $4$, and $5$.
\begin{figure} [t]
\includegraphics[width=7.5cm]{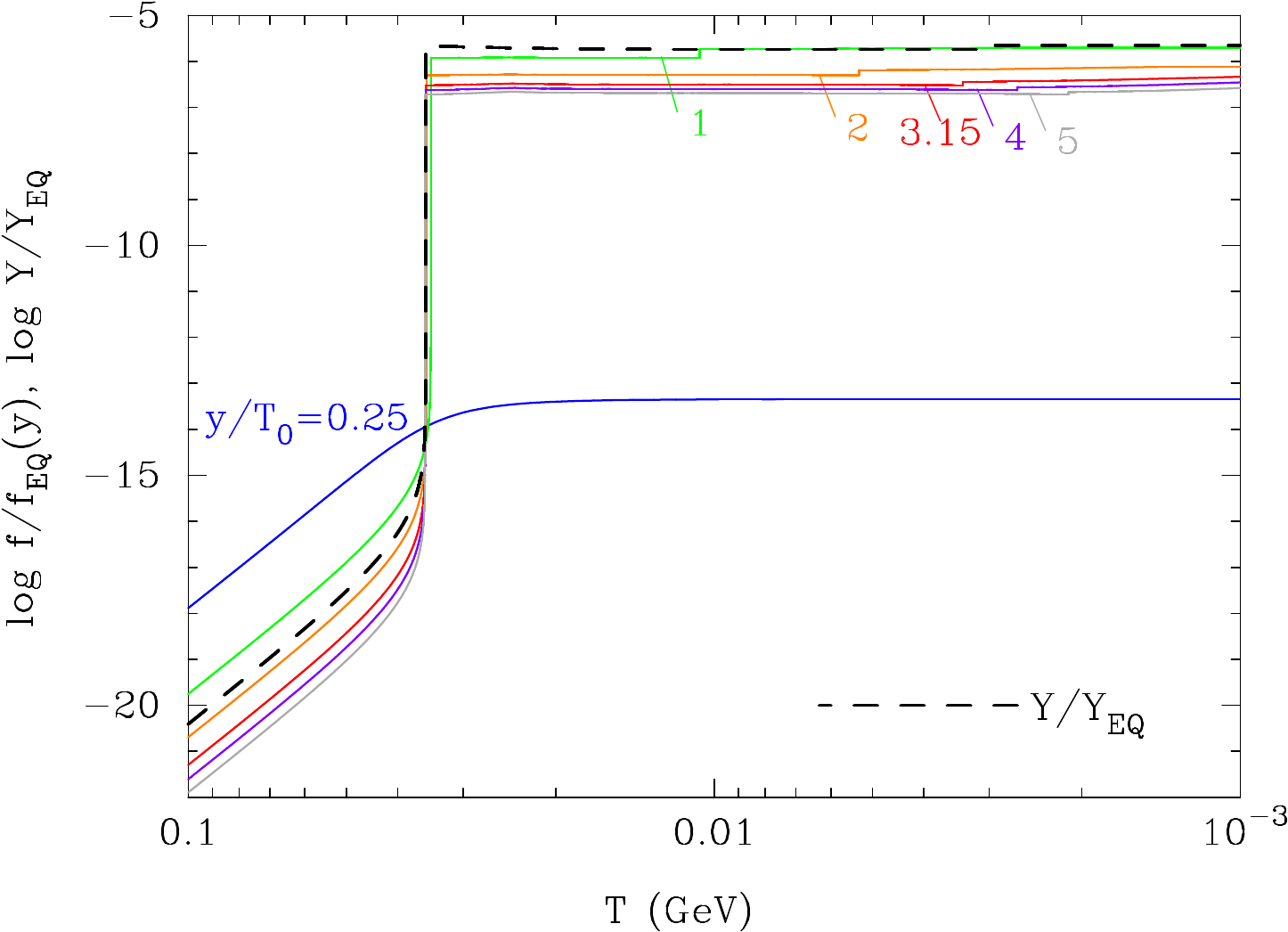}
\caption{(Color online) Temperature evolution of distribution function $f/f_{\rm EQ} (y)$ (solid lines) for $y/T_0=0.25$, $1$, $2$, $3.15$, $4$, and $5$.  The temperature evolution of the abundance $Y/Y_{\rm EQ}$ is also shown (dashed line).  The mass and the bare mixing angle of the sterile neutrino are set to $m_{\nu_{\rm s}} =1$ eV and $\theta=10^{-8}$, respectively.  The resonant energy is $E_{\rm res} =10 $ MeV.}
\label{fig_b1}
\end{figure}
At high temperature, the effective mixing angle is hindered by the matter effect [Eq. (\ref{eq_ap1})].  As the temperature decreases, the effective mixing angle increases and the distribution function increases also.  Since the effective mixing angle is smaller for larger energy $E$, the distribution function is larger for smaller $E$ or smaller $y=E_0$ values.  At $T=35.4$ MeV, the 1 + matter term in Eq. (\ref{eq_ap1}) cancels with the extra-dimensional term in the square brackets.  Therefore, the effective mixing angle becomes large for a short time resonantly.  This first resonance occurs at the temperature [cf. Eq. (\ref{eq_ap1})]
\begin{equation}
  T_{\rm res,1} \approx \left[
    \frac{\cos 2 \theta \alpha \delta m^2}{C_\alpha G_{\rm F}^2 E_{\rm res}^2}
    \right]^{1/4}.
  \label{eq_b65}
\end{equation}

The values of distribution function then suddenly increase excepting those at low energies (see the curve for $y/T_0=0.25$).  This resonance does not exist for low energies for the following reason:   When the matter term becomes smaller than the extra-dimensional term, the absolute value $|(E/E_{\rm res})|^2$ is already relatively small.  Therefore, the square brackets does not become very close to zero and the strong resonance of $\sin^2 2\tilde{\theta} \approx 1$ is never realized.

After the first resonance temperature, the second resonance occurs at a temperature which is significantly dependent on the energy $y$. One can see a slight increase in the distribution function $f(y)$ at the second resonance. In general, at this point, the matter term becomes negligible and the extra-dimensional term cancels with unity in the square brackets of Eq. (\ref{eq_ap1}).  This resonance approximately occurs at the time when the sterile neutrino energy is identical to the resonant energy $E_{\rm res}$.  The second resonance is then given by the condition [cf. Eqs. (\ref{eq_dis_4}) and (\ref{eq_dis_6})]
\begin{equation}
  T_{\rm res,2} (y) \approx T_0 \frac{E_{\rm res}}{y}  \left( \frac{g_{\ast {\rm S}0}}{g_{\ast {\rm S},{\rm res}}(y)} \right)^{1/3}.
  \label{eq_b66}
\end{equation}
The second resonant temperature becomes the smaller for the larger energies $y$. The dashed line shows the abundance ratio $Y/Y_{\rm EQ}$ calculated by solving the rate equation [Eq. (\ref{eq8})].  It is close to the ratio of the distribution function $f/f_{\rm EQ}(3.15 T_0)$, i.e., the value for the average energy of the equilibrium distribution, although a difference by a factor exists between the two lines.

Figure \ref{fig_b2} shows the effective mixing angle as a function of temperature for $y/T_0=0.25$, $1$, $2$, $3.15$, $4$, and $5$. No resonance exists for the low energy of $y/T_0=0.25$ as explained above, and there are two resonances for other energies.

\begin{figure} [t]
\includegraphics[width=7.5cm]{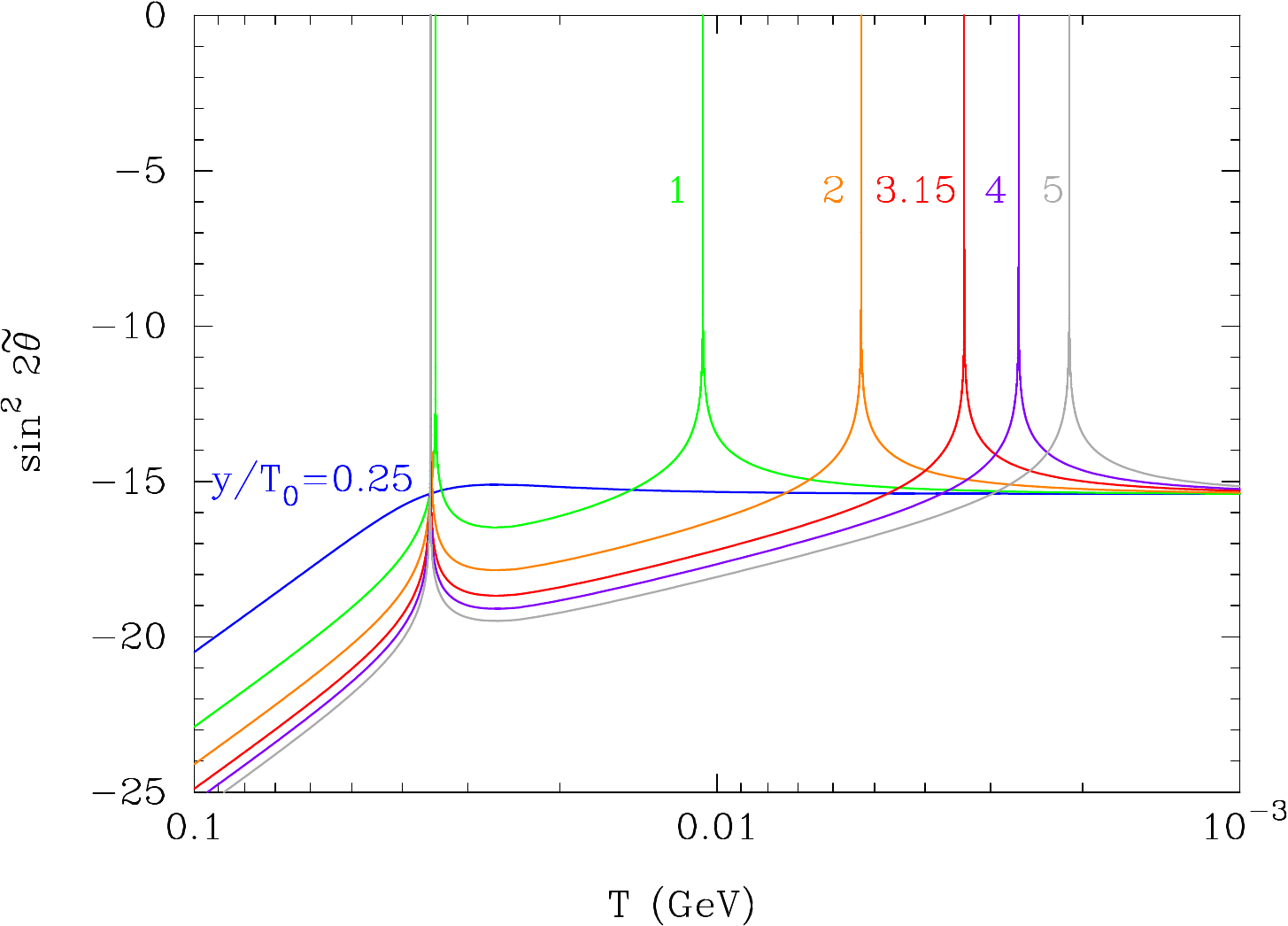}
\caption{(Color online) The effective mixing angle as a function of temperature for $y/T_0=0.25$, $1$, $2$, $3.15$, $4$, and $5$.  Adopted parameters are the same as in Fig \ref{fig_b1}.}
\label{fig_b2}
\end{figure}

Figure \ref{fig_b3} shows the distribution function of the sterile neutrino as a function of the initial energy $y=E_0$ at $T=100$, $40$, $35$, $30$, $10$, and $3$ MeV (solid lines).  At $T=100$ MeV, no resonance has come for the effective mixing angle, and the distribution function is low totally and higher for low energies (cf. Fig. \ref{fig_b1}).  At $T=40$ MeV before the first resonance, the distribution function is larger but still very small. At $T=35$ MeV during the first resonance, the distribution function is suddenly increasing. This increase occurs from larger $y$ to lower $y$. The 1st resonance occurs at $FE^2 =-1$, which is realized earlier, i.e., at higher $T$, for larger $y$ [see Eq. (\ref{eq_full8})]. At $T =30$ MeV, the distribution function is large for energies larger than $y\sim 50$ GeV.  At $T=10$ MeV, the large value of distribution function is extended to somewhat lower energy $y$, and a slight increase of the function for $y\lesssim 100$ GeV is observed. This slight increase is caused by the second resonance which occurs earlier for lower $y$ values. At $T=3$ MeV, the distribution function in the range of $y\sim[100, 350]$ GeV is larger than that of $T=10$ MeV because of the effect of the second resonance.

\begin{figure} [t]
\includegraphics[width=7.5cm]{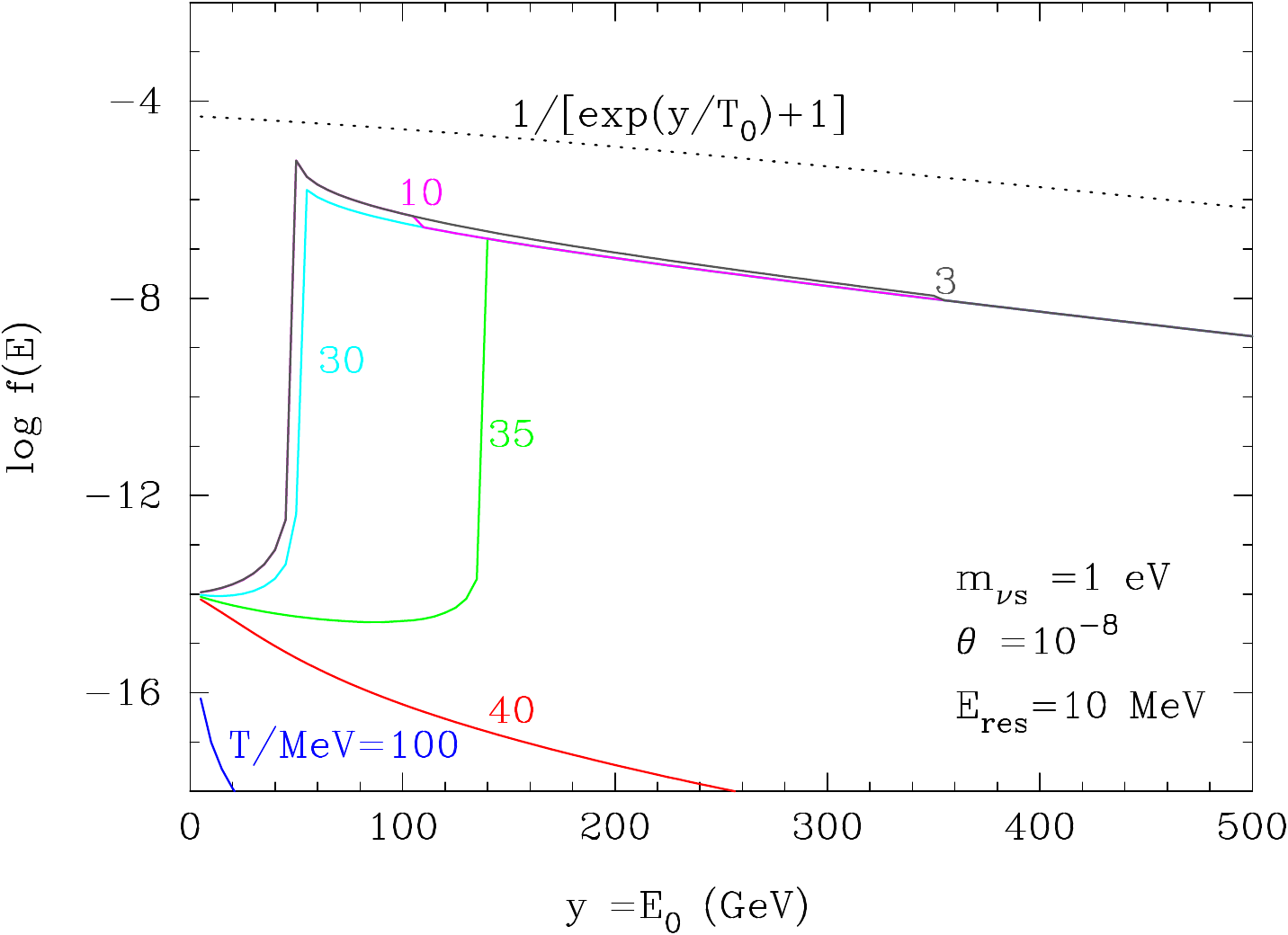}
\caption{(Color online) The distribution function of the sterile neutrino as a function of $y=E_0$ at $T=100$, $40$, $35$, $30$, $10$, and $3$ MeV (solid lines).  Adopted parameters are the same as in Fig \ref{fig_b1}. The dotted line is the equilibrium function at the initial temperature $T_0$ that is normalized arbitrarily.}
\label{fig_b3}
\end{figure}

The dotted line is the equilibrium function at the initial temperature $T_0$ that is normalized arbitrarily. As seen from the equilibrium function and the last distribution function at $T=3$ MeV, the real distribution function is different from the equilibrium spectrum. Main differences are (1) the cutoff energy below which the distribution function is very small because of no resonance, and (2) a different dependence of the function on energy. The production rate of sterile neutrino is larger for smaller energy of the sterile neutrino. Therefore, the increase of distribution function at the resonance is larger for smaller energies (see Fig. \ref{fig_b1}).  As a result, the final distribution function for low energies is enhanced with respect to the equilibrium spectrum.

We derive the final energy density of sterile neutrino $\rho_{\nu_{\rm s}} =1.3 \times 10^{-19}$ GeV$^4$ from the integration of Boltzmann equation. The approximate energy density from the integration of rate equation is $\rho_{\nu_{\rm s}} =6.4 \times 10^{-19}$ GeV$^4$. It is then found that the use of the rate equation gives a rough estimation of the sterile neutrino energy density although there are significant differences in spectra and the total number densities from values of the calculation of an exact Boltzmann equation.
\end{widetext}


\end{document}